\documentclass{elsarticle}

\usepackage{epsfig}

\usepackage{amssymb}
\usepackage{newtxtext,newtxmath,amsmath,natbib}

\usepackage{color}
\usepackage{lscape}
\usepackage{caption}

\usepackage{lineno}

\journal{Earth and Planetary Science Letters}

\begin{document}

\begin{frontmatter}

\title{Subduction and atmospheric escape of Earth's seawater constrained by hydrogen isotopes}

\author[label1]{Hiroyuki Kurokawa\corref{cor1}}
\address[label1]{Earth-Life Science Institute, Tokyo Institute of Technology, 2-12-1 Ookayama, Meguro, Tokyo 152-8550, Japan}

\cortext[cor1]{Corresponding author}

\ead{hiro.kurokawa@elsi.jp}

\author[label1]{Julien Foriel}

\author[label1]{Matthieu Laneuville}

\author[label1]{Christine Houser}

\author[label1]{Tomohiro Usui}

\begin{abstract}
The hydrogen isotopic (D/H) ratio reflects the global cycling and evolution of water on Earth as it fractionates through planetary processes. 
We model the water cycle taking seafloor hydrothermal alteration, chemical alteration of continental crust, slab subduction, hydrogen escape from the early Earth, and degassing at mid-ocean ridges, hot spots, and arcs into account.
The differences in D/H ratios between present-day oceans, oceanic and continental crust, and mantle are thought to reflect isotopic fractionation through seafloor alteration, chemical alteration, and slab dehydration.
\textcolor{black}{However}, if the speed of plate tectonics has been nearly constant through out Earth's history, the degassing and regassing rates are too small to reach the present-day D/H ratios.
We show that (a) hydrogen escape from reduced early atmosphere, (b) secular net regassing, or (c) faster plate tectonics on early Earth is needed to reproduce the present-day D/H ratios of the water reservoirs.
The low D/H ratio of \textcolor{black}{Archean} seawater at 3.8 Ga \textcolor{black}{has previously been interpreted as a signature of (a) hydrogen escape, but we find it} can also be explained either by (b) secular net regassing or by (c) faster plate tectonics on early Earth.
The rates of hydrogen escape from early Earth and secular regassing on present-day Earth are constrained to be lower than \textcolor{black}{$2.1\times 10^{11}$ kg/yr} and $3.9\times 10^{11}$ kg/yr. 
Consequently, the volume of water in the present-day mantle could result entirely from the regassing through Earth's history. 
In that case, the volume of initial oceans could be 2 to 3 times larger than that of current Earth.
We suggest that, in addition to the D/H ratio of Archean seawater, identifying the  D/H ratios of both seawater and mantle \textcolor{black}{throughout Earth's history} would allow to distinguish these evolutionary scenarios.
\end{abstract}


\end{frontmatter}


\section{Introduction}
\label{Introduction}

Water plays a critical role in controlling the physical and chemical evolution on Earth through atmosphere-ocean-continent interactions which control the atmospheric composition, the climate through the carbon cycle, the subduction of water, and possibly even the emergence and evolution of life \citep[e.g.,][]{Gaillard+2011,Walker1977,Honing+2014,Honing+2016,Dohm+Maruyama2015}.
Furthermore, the abundance of water in Earth's interior influences the mantle melting, rheology, and style of convection \citep[e.g.,][]{Hirschmann2006,Karato+Jung2003,Mei+Kohlstedt2000}.

The abundance of water on the surface and in the interior is controlled by the deep water cycle between the oceans and mantle, and loss caused by hydrogen escape throughout Earth's history.
Slab subduction transports water as bound and pore water in metamorphic rocks and sediments that originate from hydrothermal alteration of the seafloor and chemical alteration of continental crust \citep[e.g.,][]{Jarrard+2003,Bodnar+2013,Honing+2014,Honing+2016}.
While the majority of the subducted water returns to the oceans directly by updip transport and indirectly by arc volcanism, some trace amounts of water may remain in the mantle.
Water can return from the mantle to the exosphere (here defined as the atmosphere and hydrosphere) by regassing at mid-ocean ridges and ocean islands.
Although the photolysis of water has a negligible effect on the loss of hydrogen from the atmosphere today, the hydrogen escape through the photolysis of methane in the reduced early atmosphere before the great oxidation event (GOE) at 2.5 Ga could have a more significant impact \citep{Catling+2001}.

Despite its importance to control the water budget, the balance between the degassing and regassing as well as the early hydrogen escape flux is poorly understood.
\textcolor{black}{We note that the term \textquotedblleft regassing\textquotedblright \ means the water transport to the mantle as bound and pore water in metamorphic rocks and sediments.}
Net regassing from the oceans to the mantle has been suggested from the geochemical estimates \citep{Ito+1983}.
The continental freeboard is proposed to be nearly constant from the end of the Archean \citep[e.g.,][]{Schubert+Reymer1985} and is interpreted as the degassing and regassing rates almost achieving balance \citep[e.g.,][]{Lecuyer+1998,Parai+Mukhopadhyay2012}.
However, \citet{Korenaga+2017} recently argued that the relative buoyancy of continental lithosphere with respect to oceanic lithosphere was higher in the past, which requires a larger volume of oceanic water at the time to keep continental freeboard constant.
Hydrogen escape on early Earth is even more poorly constrained as the atmospheric composition at that time is not well known.

Hydrogen isotope (D/H) ratio has been used to constrain the global cycle and loss of water on Earth, as it fractionates through planetary processes. 
Earth's mantle is known to have ${\rm \delta D} = -80$\textperthousand \ to -60 \textperthousand \ (${\rm \delta D = [(D/H)_{sample}/(D/H)_{reference} -1] \times 10^3}$, where the reference is the standard mean ocean water, hereafter SMOW), which is lower than that of today's oceans (defined here as the total hydrosphere) \citep{Kyser+ONeil1984,Clog+2013}.
The low mantle ${\rm \delta D}$ value has been considered to suggest that the mantle became isolated from the oceans through geologic time \citep{Kyser+ONeil1984}, or that water in the mantle has been isotopically fractionated from \textcolor{black}{ the source seawater because of seafloor alteration and slab dehydration processes \citep{Lecuyer+1998,Shaw+2008,Shaw+2012}.}
\textcolor{black}{Isotopic analysis of Archean minerals and rocks has found that Archean seawater has a ${\rm \delta D}$ value lower than that of present-day oceans, which has been interpreted as a signature of water loss caused by the hydrogen escape \citep{Hren+2009,Pope+2012}.}

\textcolor{black}{While D/H ratio has been widely utilized to constrain those processes of the water cycle and loss, there is no comprehensive model of the D/H evolution which involves all relevant processes.
Previous studies considered the degassing and regassing \citep{Lecuyer+1998,Shaw+2008} or the hydrogen escape \citep{Pope+2012} only.
In addition, the D/H ratios of the mantle and Archean seawater have been considered separately to constrain these different processes.
Because all water reservoirs are coupled to each other, all these processes and D/H constraints should be considered simultaneously, which is the aim of this study.
}

We model the global water cycle taking seafloor hydrothermal alteration, chemical alteration of continental crust, slab subduction, atmospheric escape, and degassing at mid-ocean ridges, hot spots, and arcs into account.
The model calculations are compared with the D/H ratios of water in different reservoirs on present-day Earth and of Archean seawater to constrain the rates of hydrogen escape from early Earth and of secular regassing on present-day Earth.
Section \ref{Model} presents the model.
Section \ref{Results} shows the results.
The implications for the evolution of water on Earth are discussed in Section \ref{Discussion}.
We conclude in Section \ref{Conclusion}.

\section{Methods}
\label{Model}

\subsection{Model}
\label{sub:model}

\begin{figure}
    \centering
    \includegraphics[width=10cm]{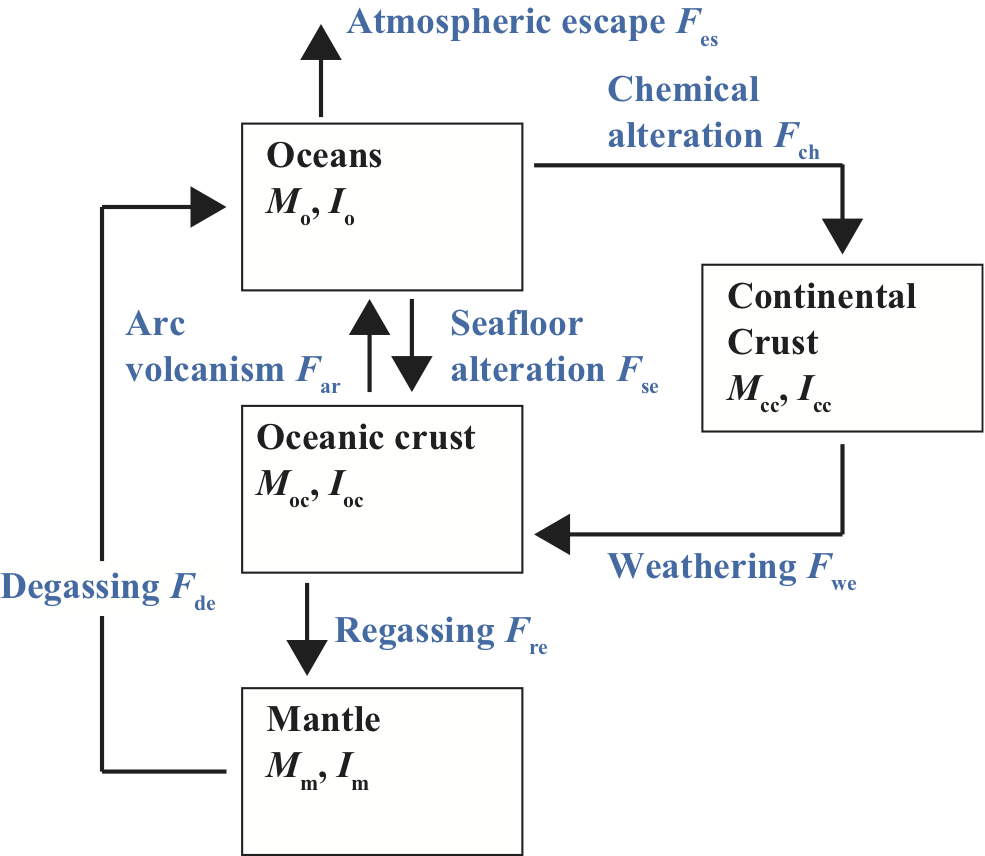}
    \caption{Schematic view of Earth's water cycle in our model.}
    \label{fig:model}
\end{figure}

We constructed a global water cycle model taking the D/H compositions into account. 
Four reservoirs were considered in our model: the oceans, continental crust, oceanic crust, and mantle.
These reservoirs exchange water through seafloor hydrothermal alteration, chemical alteration of continents, slab subduction, and degassing at mid-ocean ridges, hot spots, and arcs (Figure \ref{fig:model}).
Hydrogen loss to space induced by photolysis of methane was considered to have occurred on early Earth before the GOE at 2.5 Ga. 
The oceans in our model include water in small reservoirs that exchange water with oceans over short timescales: atmosphere, biosphere, surface water, ground water, and glaciers/polar ice, see Table \ref{Table_reservoirs}.
Hereafter the oceans with the small reservoirs are referred to as \textquotedblleft bulk oceans\textquotedblright\ when we would like to distinguish them from seawater without small reservoirs.
Exploring the possible ranges of fluxes, we constrain the water cycle and discuss the implications for the evolution of water on Earth.

Evolution of the masses and D/H ratios of water in each reservoir was calculated by using the following equations, 
\begin{equation}
    {\rm \frac{d{\it M}_{i}}{d{\it t}} = \sum_{sources} {\it F}_{k} - \sum_{sinks} {\it F}_{k} } \label{eq:Mi}
\end{equation}
\begin{equation}
    {\rm \frac{d}{d{\it t}} ({\it M}_{i}{\it I}_{i}) = \sum_{sources} {\it F}_{k}{\it f}_{k}{\it I}_{i'} - \sum_{sinks} {\it F}_{k}{\it f}_{k}{\it I}_{i}} \label{eq:MiIi}
\end{equation}
\begin{equation}
{\rm \frac{{\it f}_{re}}{{\it f}_{ar}} =  {\it f}_{dehy}  } \label{eq:fdehy1}
\end{equation}
\begin{equation}
{\rm \frac{{\it F}_{ar}}{{\it F}_{ar}+{\it F}_{re}} {\it f}_{ar} + \frac{{\it F}_{re}}{{\it F}_{ar}+{\it F}_{re}} {\it f}_{re} = 1} \label{eq:fdehy2}
\end{equation}
where $M_{\rm i}$ and $I_{\rm i}$ are the mass and D/H ratio of water in the reservoir i, and $F_{\rm k}$ and $f_{\rm k}$ are the flux of the process k and its fractionation factor.
The sources and sinks for each reservoir are described in Figure \ref{fig:model}.
Hereafter subscripts i=o, cc, oc, and m denote the oceans, continental crust, oceanic crust, and mantle, respectively.
The subscript i' in Equation \ref{eq:MiIi} denotes a reservoir other than i.
Subscripts k = ch, ar, se, de, es, we, and re denote the chemical alteration, arc volcanism, seafloor alteration, degassing, atmospheric escape, weathering, and regassing, respectively.
Equations \ref{eq:fdehy1} and \ref{eq:fdehy2} give $f_{\rm ar}$ and $f_{\rm de}$ by considering the dehydration-induced fractionation and mass balance.
Assuming ${\rm d}/{\rm d}t = 0$ in Equation \ref{eq:MiIi} gives a steady state in D/H, which is useful to understand the \textcolor{black}{numerical} results (\ref{appendix-sec1}).

Our model assumed that the fluxes depend on the masses of water in the reservoirs and time as follows: 
\begin{equation}
    F_{\rm ch} = F_{\rm ch}^0 \frac{A_{\rm c}(t)}{A_{\rm c}^0} \label{eq:Fch}
\end{equation}
\begin{equation}
    F_{\rm se} = F_{\rm se}^0 \times f(t) \label{eq:Fse}
\end{equation}
\begin{equation}
    F_{\rm ar} = F_{\rm ar}^0 \frac{M_{\rm oc}(t)}{M_{\rm oc}^0} \times f(t) \label{eq:Far}
\end{equation}
\begin{equation}
    F_{\rm de} = F_{\rm de}^0 \frac{M_{\rm m}(t)}{M_{\rm m}^0} \times f(t)^{\frac{1}{2}} \label{eq:Fde}
\end{equation}
\begin{equation}
    F_{\rm we} = F_{\rm we}^0 \frac{M_{\rm cc}(t)}{M_{\rm cc}^0} \label{eq:Fwe}
\end{equation}
\begin{equation}
    F_{\rm re} = F_{\rm re}^0 \frac{M_{\rm oc}(t)}{M_{\rm oc}^0} \times f(t) \label{eq:Fre}
\end{equation}
where superscript 0 denotes the reference (present-day) values, $A_{\rm c}$ is the continental area, and $f(t)$ is the speed of plate tectonics scaled to that on present-day Earth, which accounts for the change of mantle-convection speed due to the cooling suggested by conventional thermal-evolution models \citep[e.g.,][]{Stevenson+1983,Honing+2016}.
The difference in the dependence on $f(t)$ between Equations \ref{eq:Fse}, \ref{eq:Far}, \ref{eq:Fre} and \ref{eq:Fde} originates from the boundary-layer model \citep{Honing+2016}.

The continental area $A_{\rm c}$ as a function of time is given by \citep{Mclennan+Taylor1982},
\begin{equation}
\frac{A_{\rm c}}{A_{\rm c}^0} = \begin{cases}
    0.1875\times \Bigl( \frac{t}{\rm Gyr} \Bigr) & (t < 0.8\ {\rm Gyr}) \\
    0.15 + 0.929\times \Bigl( \frac{t}{\rm Gyr}-0.8 \Bigr) & (0.8\ {\rm Gyr} < t < 1.5\ {\rm Gyr}) \\
    0.8 + 0.0667\times \Bigl( \frac{t}{\rm Gyr}-1.5 \Bigr) & (1.5\ {\rm Gyr} < t) \label{eq:Ac}
  \end{cases}
\end{equation}
\textcolor{black}{We note that, though the evolution of the continental coverage is controversial, our results are shown to depend only weakly on the choice of the continental growth model (subsection \ref{ss:continents}).
}

The plate speed scaled by the present-day value $f(t)$ is given by, 
\begin{equation}
    f(t) = 10^{-(t/4.5{\rm Gyr}-1)} \label{eq:ft}
\end{equation}
Equation \ref{eq:ft} reflects the model of \citet{Honing+2016}, which argued that the speed of subduction was $\sim 2-10$ times faster when the potential temperature of the mantle was $\sim 100-300$ K hotter than today \citep{Herzberg+2010}.
In contrast, a recent model based on the energetics of plate-tectonic mantle convection proposed that the tectonic speed and surface heat flux have been nearly constant throughout Earth's history \citep{Korenaga2003,Korenaga+2017}, which implies $f(t) = 1$.
We considered both the former and latter cases.
Hereafter the two thermal evolution models are referred to \textcolor{black}{as} the faster and slower plate tectonics (PT) models, respectively.
Models assuming time-independent (constant) fluxes were also explored to show the basic properties of the evolution of D/H (\ref{appendix-sec2}).

\subsection{Parameters}
\label{model:parameters}

\begin{table}[]
    \centering
    \begin{tabular}{lr}
    \hline \hline
    Present-day water fluxes [$10^{11}\ {\rm kg/yr}$] & \\
    \hline
    $F_{\rm de}^0$ & 1.0 \\
    $F_{\rm se}^0$ & 10 \\
    $F_{\rm ch}^0$ & 1.5 \\
    $F_{\rm we}^0$ & 1.0 \\
    $F_{\rm re}^0$ & $F_{\rm de}^0+F_{\rm re,net}$ \\
    $F_{\rm ar}^0$ & $F_{\rm se}^0+F_{\rm we}^0-F_{\rm re}^0$ \\
    \hline \hline
    Fractionation factors & \\
    \hline
    $10^3 {\rm ln} f_{\rm de}$ & 0 (standard), 10 \\
    $10^3 {\rm ln} f'_{\rm se}$ & -30 \\
    $10^3 {\rm ln} f'_{\rm ch}$ & -80 \\
    $10^3 {\rm ln} f_{\rm we}$ & 0 \\
    $10^3 {\rm ln} f_{\rm re}$ & Equations \ref{eq:fdehy1} and \ref{eq:fdehy2} \\
    $10^3 {\rm ln} f_{\rm ar}$ & Equations \ref{eq:fdehy1} and \ref{eq:fdehy2} \\
    $10^3 {\rm ln} f_{\rm dehy}$ & -40 (standard), -23 \\
    $10^3 {\rm ln} f_{\rm es}$ & -150 \\
    \hline \hline
    Present-day water masses [ocean] & \\
    \hline
    $M_{\rm o}^0$ & 1 \\
    $M_{\rm cc}^0$ & 0.2 \\
    $M_{\rm oc}^0$ & 0.1 \\
    $M_{\rm m}^0$ & 1 or >1 \\
    \hline \hline
    Initial water masses & \\
    \hline
    $M_{\rm o}^{\rm i}$ & $M_{\rm o}^0$ + $M_{\rm cc}^0$ + $M_{\rm es}$ + $M_{\rm re}$ \\
    $M_{\rm cc}^{\rm i}$ & 0 ocean \\
    $M_{\rm oc}^{\rm i}$ & $M_{\rm oc}^0$ \\
    $M_{\rm m}^{\rm i}$ & $M_{\rm m}^0$ - $M_{\rm re}$ \\
    \hline
    \end{tabular}
    \caption{Summary of parameters and initial conditions (see Supplementary text S1-4 for references). 1 ocean = 1.4 $\times$ 10$^{21}$ kg. An apostrophe denotes the fractionation from the water in oceans before the correction by adding small reservoirs (Supplementary text S3).}
    \label{tab:model}
\end{table}

Our assumptions on parameters are summarized in Table \ref{tab:model} (Supplementary text S1).
We assumed $M_{\rm o}^0 = 1$ ocean, $M_{\rm cc}^0 = 0.2$ oceans, and $M_{\rm oc}^0 = 0.1$ oceans (1 ocean = 1.4 $\times$ 10$^{21}$ kg).
Most of our models assumed $M_{\rm m}^0 = 1$ ocean, but some models showed the integrated regassing to be larger than 1 ocean.
In these cases, we increased $M_{\rm m}^0$ \textcolor{black}{iteratively} until the model results in $M_{\rm m}$ at present which agrees with the assumed $M_{\rm m}^0$.

The present-day net-regassing flux $F_{\rm re,net}$ $(= F_{\rm re}^0-F_{\rm de}^0)$ and escape flux before 2.5 Ga $F_{\rm es}$ are treated as independent parameters.
We assumed $F_{\rm de}^0 = 1.0 \times 10^{11}\ {\rm kg/yr}$, $F_{\rm se}^0 = 10 \times 10^{11}\ {\rm kg/yr}$, $F_{\rm ch}^0 = 1.5 \times 10^{11}\ {\rm kg/yr}$, $F_{\rm we}^0 = 1.0 \times 10^{11}\ {\rm kg/yr}$, $F_{\rm re}^0 = F_{\rm de}^0+F_{\rm re,net}^0$, and $F_{\rm ar}^0 = F_{\rm se}^0+F_{\rm we}^0-F_{\rm re}^0$, respectively (Supplementary text S2).
Models assuming different values of the fluxes \textcolor{black}{were} also explored to show the basic behavior of the system (\ref{appendix-sec2}).

Our standard model assumed the values of fractionation factors as summarized in Table \ref{tab:model}.
Considering the uncertainty of fractionation factors (Supplementary text S3), we also examined the cases where different values were assumed.

\textcolor{black}{
While our model assumed that the majority of Earth's water had been delivered before the solidification of magma oceans, part of water might have been delivered by the late accretion of comets.
In order to investigate the influence of the possible late accretion of comets, we also explored models where the cometary delivery was implemented by an input of $0.01$ ocean water with $\delta {\rm D} = 1000$\textperthousand \ at $4.1$ Ga (Supplementary text S5).
}

\subsection{Initial conditions}
\label{model:initial}

The initial conditions of water volumes were assumed as follows (Table \ref{tab:model} and Supplementary text S4).
Because there was no continental crust initially (Equation \ref{eq:Ac}), $M_{\rm cc}^{\rm i}= 0$ oceans was assumed.
The water in sediments on present-day Earth was assumed to be initially partitioned into the oceans.
We further assumed that the integrated water lost by escape $M_{\rm es}$ originated from the oceans.
Therefore, the initial water in oceans was given by $M_{\rm o}^{\rm i} = M_{\rm o}^0 + M_{\rm cc}^0 + M_{\rm es} + M_{\rm re}$, where $M_{\rm re}$ is the integrated net regassing.
The regassed water $M_{\rm re}$ is not given {\it a priori}, but it was given by iteration.
The initial mantle water was given by $M_{\rm m}^{\rm i} = M_{\rm m}^0 - M_{\rm re}$.

\textcolor{black}{
We assumed that all the reservoirs have the same ${\rm \delta D}$ and iteratively change the value to obtain the present-day oceanic ${\rm \delta D}$ value which equals the SMOW (Supplementary text S4). 
We note that the initial ${\rm \delta D}$ values obtained by this procedure were within the range of carbonaceous chondrites: ${\rm \delta D} = -200$\textperthousand \ to 300\textperthousand \ \citep{Lecuyer+1998}.
}

\subsection{Constraints from D/H}
\label{model:D/H}

\begin{table}[]
    \centering
    \begin{tabular}{ lrr }
 \hline \hline
 Reservoir & Amount of water (ocean) & ${\rm \delta D}$ (\textperthousand) \\
 \hline \hline
 Oceans (including small reservoirs) & 1.0 & -12 to -7.1 \\ 
 \hline
 \ \ Oceans & 0.98\ \footnotemark[1] & 0\ \footnotemark[1] \\
 \ \ Atmosphere & $9.3\times 10^{-6}$\ \footnotemark[1] & -70 to +10 \footnotemark[2] \\
 \ \ Biosphere & $3.4\times 10^{-5}$\ \footnotemark[1] & -130 to -70\ \footnotemark[3] \\
 \ \ Surface water & $1.5\times 10^{-4}$\ \footnotemark[1] & -300 to +10\ \footnotemark[4] \\
 \ \ Ground water & $7.5\times 10^{-3}$\ \footnotemark[1] & -300 to +10\ \footnotemark[4] \\
 \ \ Glaciers/Polar ice & $2.4\times 10^{-2}$\ \footnotemark[1] & -400 to -300\ \footnotemark[3] \\
 \hline
 Oceanic crust & 0.10\ \footnotemark[1] & -50 to -30\ \footnotemark[6] \\ 
 \hline
 Continental crust & 0.20\ \footnotemark[1] & -100 to -60 \footnotemark[3] \\ 
 \hline
 Mantle & 1.0\ \footnotemark[4] & -80 to -60\  \footnotemark[7] \\ 
 \hline
\end{tabular}
    \caption{Sizes and D/H ratios of water reservoirs on present-day Earth. 1 ocean = $1.4\times 10^{21}$ kg. 1: \citet{Bodnar+2013} and references therein. 2: \textquotedblleft Volumetrically most important meteoric waters" from \citet{Sheppard1986}. 3: \citet{Lecuyer+1998} and references therein. 4: \citet{Pope+2012} and references therein. 5: \citet{Korenaga+2017} and references therein. 6: \citet{Lecuyer+1998,Shaw+2008} and references therein. 7: \citet{Kyser+ONeil1984,Clog+2013}.}
    \label{Table_reservoirs} 
\end{table}

Models are considered successful if they satisfy the constraints on the D/H ratios of present-day reservoirs (Table \ref{Table_reservoirs}) and of the Archean seawater.
Present-day Earth's seawater have ${\rm \delta D} = 0$\textperthousand \ by definition.
Summing up water in oceans and small reservoirs (atmosphere, biosphere, surface water, groundwater, and glaciers) leads to the bulk oceanic ${\rm \delta D} = -12$\textperthousand \ to -7.1\textperthousand.
The gap from SMOW mostly originated from the contribution of low ${\rm \delta D}$ glaciers and polar ice.
Sedimentary rocks on continental crust and metamorphic rocks on oceanic crust have ${\rm \delta D} = -100$\textperthousand \ to -60\textperthousand \ and ${\rm \delta D} = -50$\textperthousand \ to -30\textperthousand, respectively \citep{Lecuyer+1998,Shaw+2008}.
Earth's mantle has ${\rm \delta D} = -80$\textperthousand \ to -60 \textperthousand \ \citep{Kyser+ONeil1984,Clog+2013}.

\textcolor{black}{
The D/H ratio of paleo-seawater has been estimated from isotopic analysis of minerals and rocks which interacted with seawater \citep{Wenner+Taylor1974,Lecuyer+1996,Kyser+1999,Hren+2009,Pope+2012}.
We adopted ${\rm \delta D} = -25 \pm 5$\textperthousand \ proposed by \citet{Pope+2012} as the constraint on 3.8 Ga seawater, because they have derived this value by combining hydrogen and oxygen isotope measurements of serpentine samples, in which primitive isotopic signatures have been well preserved. 
We will discuss other data sets in subsection \ref{ss:datasets}.
}
We assumed that the range of ${\rm \delta D}$ values between those of the bulk ocean and the ocean (seawater) should match the evaluated ${\rm \delta D}$ of the Archean seawater.
Assuming constant ${\rm ln}f'$ corresponds to the case where the mass ratio of glaciers/polar ice to bulk oceans has been constant through time.
Because there is no evidence of glaciation before approximately 2.9 Ga \citep{Young+1998}, considering the range of ${\rm \delta D}$ is a conservative assumption.

\section{Results}
\label{Results}

\begin{figure}
    \centering
     \includegraphics[width=12cm]{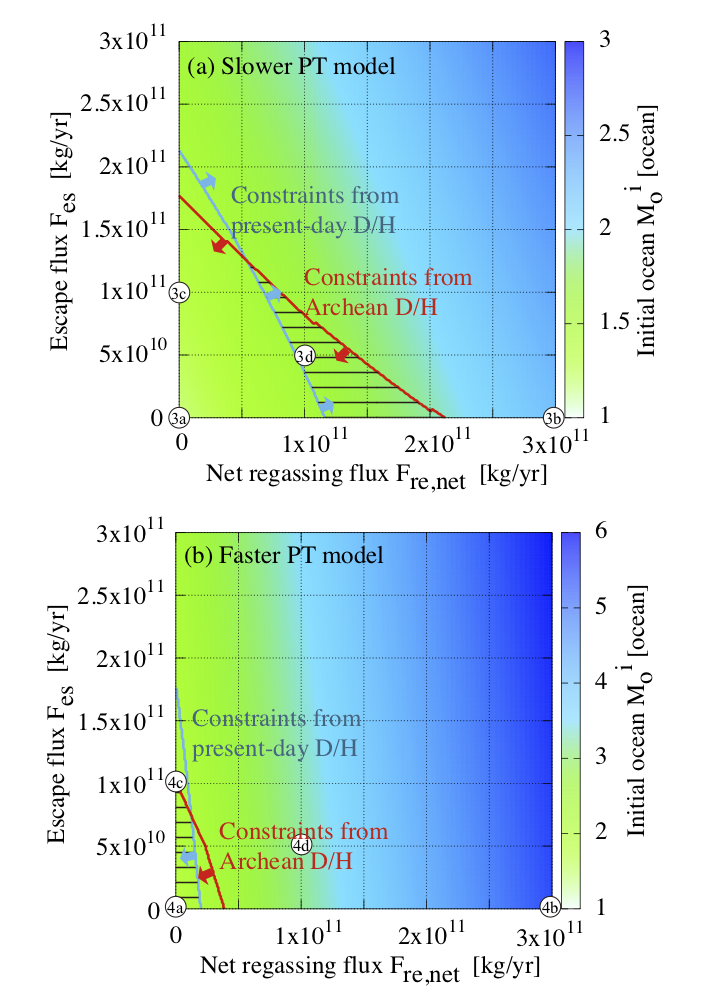}
    \caption{Range of $F_{\rm res,net}$ and $F_{\rm es}$ where the constraints on D/H are satisfied (the hatched areas). Results for (a) slower plate tectonics (PT) model and (b) faster PT model are shown. The present-day D/H ratios of the water reservoirs were reproduced above and below the sky-blue line for (a) and (b), respectively. The D/H of the Archean seawater was reproduced below the red line. Color contour denotes $M_{\rm o}^{\rm i}$. Marks in figures correspond to the parameter sets shown in Figures \ref{fig:Ts} and \ref{fig:Tf}.}
    \label{fig:FreFes}
\end{figure}

\begin{figure}
    \centering
    \includegraphics[width=12cm]{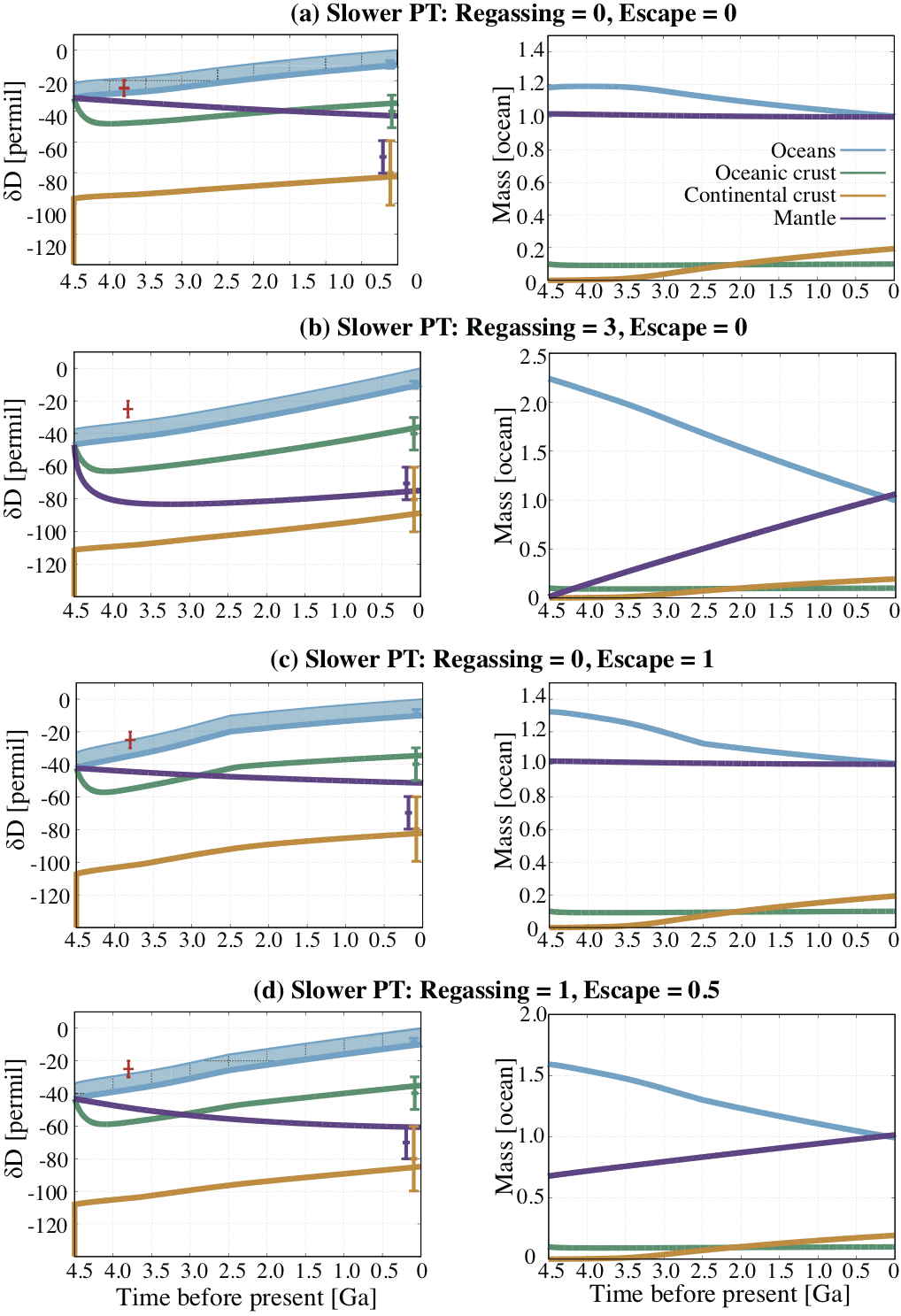}
    \caption{Time evolution of the points labelled 3a-d from the slower PT models in Figure \ref{fig:FreFes}a. Left: ${\rm \delta D}$ of oceans (thin cyan lines), bulk oceans (thick cyan lines), oceanic crust (green lines), continental crust (yellow lines), and mantle (purple lines) as a function of time. Data points are ${\rm \delta D}$ values of reservoirs on present-day Earth (subsection \ref{model:D/H} and Table \ref{Table_reservoirs}) and 3.8 Ga seawater \citep{Pope+2012}. The shaded range denotes the possible range of oceanic ${\rm \delta D}$ (see subsection \ref{model:D/H}). Right: masses of water in bulk oceans (cyan lines), oceanic crust (green lines), continental crust (yellow lines), and mantle (purple lines) as a function of time. }
    \label{fig:Ts}
\end{figure}

\begin{figure}
    \centering
    \includegraphics[width=12cm]{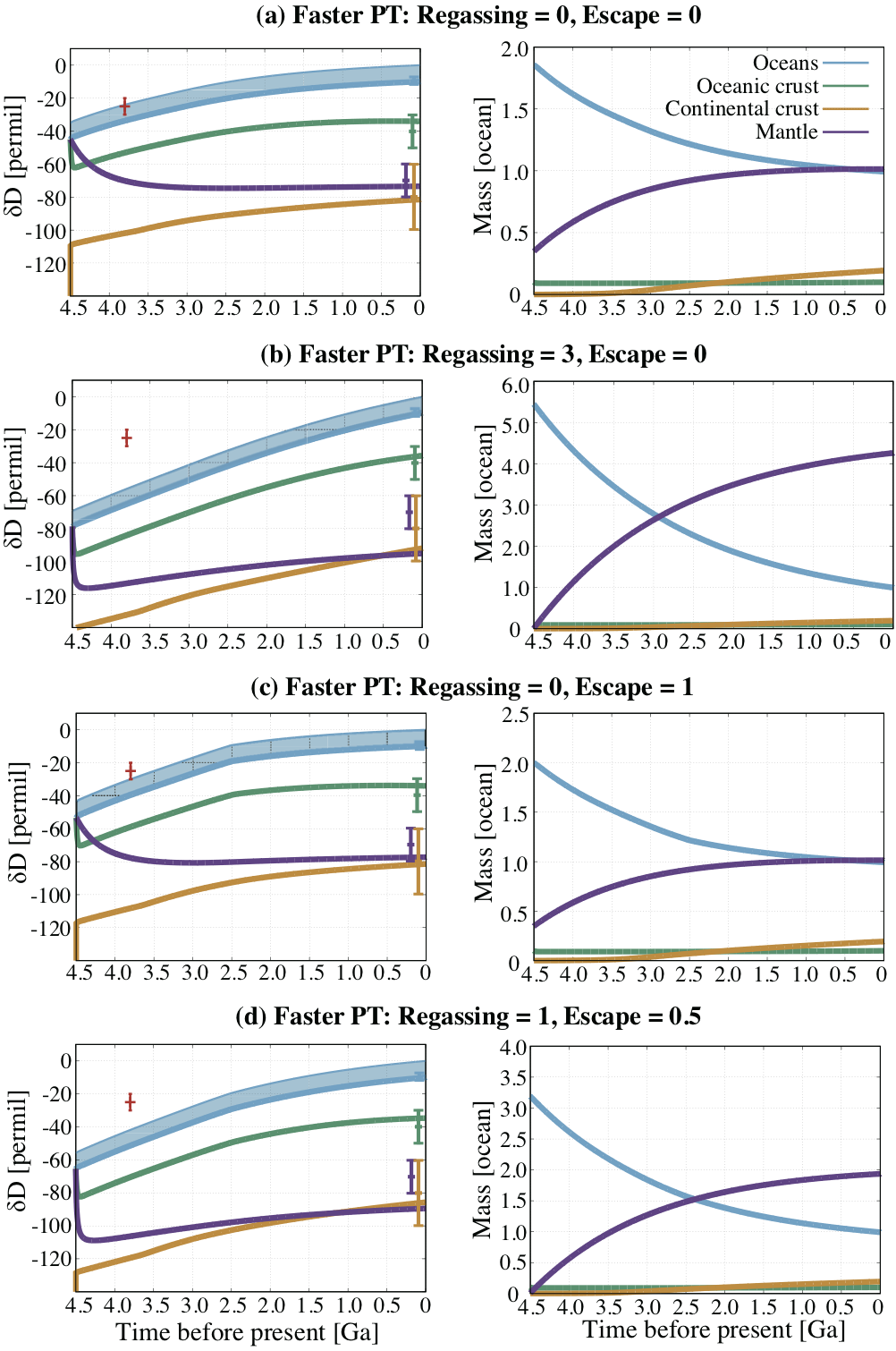}
    \caption{The same as Figure \ref{fig:Ts}, but the time evolution for the points labelled 4a-d from the faster PT models in Figure \ref{fig:FreFes}b are shown.}
    \label{fig:Tf}
\end{figure}

\begin{figure}
    \centering
    \includegraphics[width=12cm]{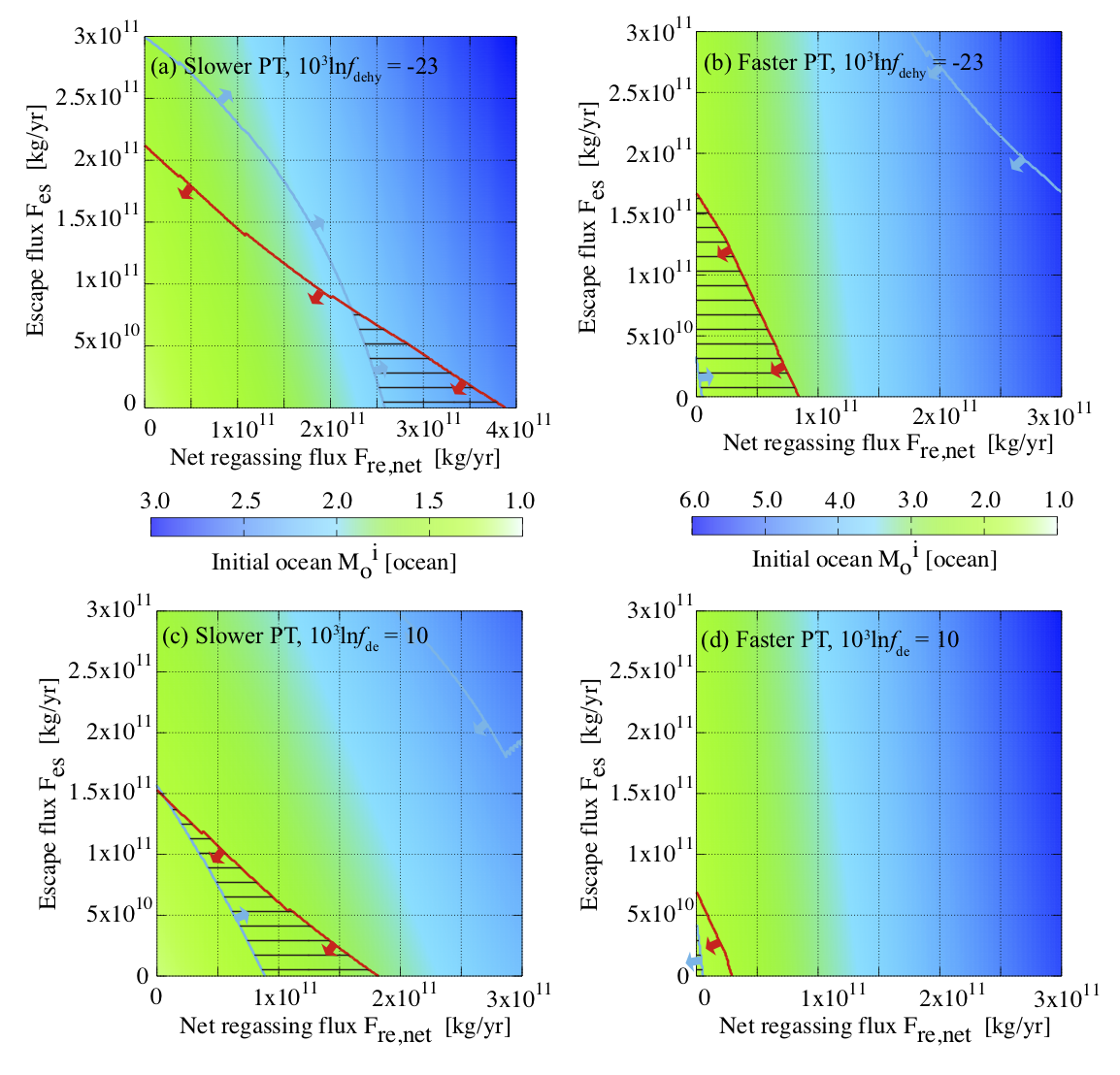}
    \caption{
    The same as Figure \ref{fig:FreFes}, but different values of fractionation factors are assumed (Table \ref{tab:model}). The cases for (a,b) $10^3 {\rm ln} f_{\rm dehy} = -23$ and (c,d) $10^3 {\rm ln} f_{\rm de} = 10$ are shown.}
    \label{fig:FreFes_fdehyfde}
\end{figure}

\begin{figure}
    \centering
    \includegraphics[width=14cm]{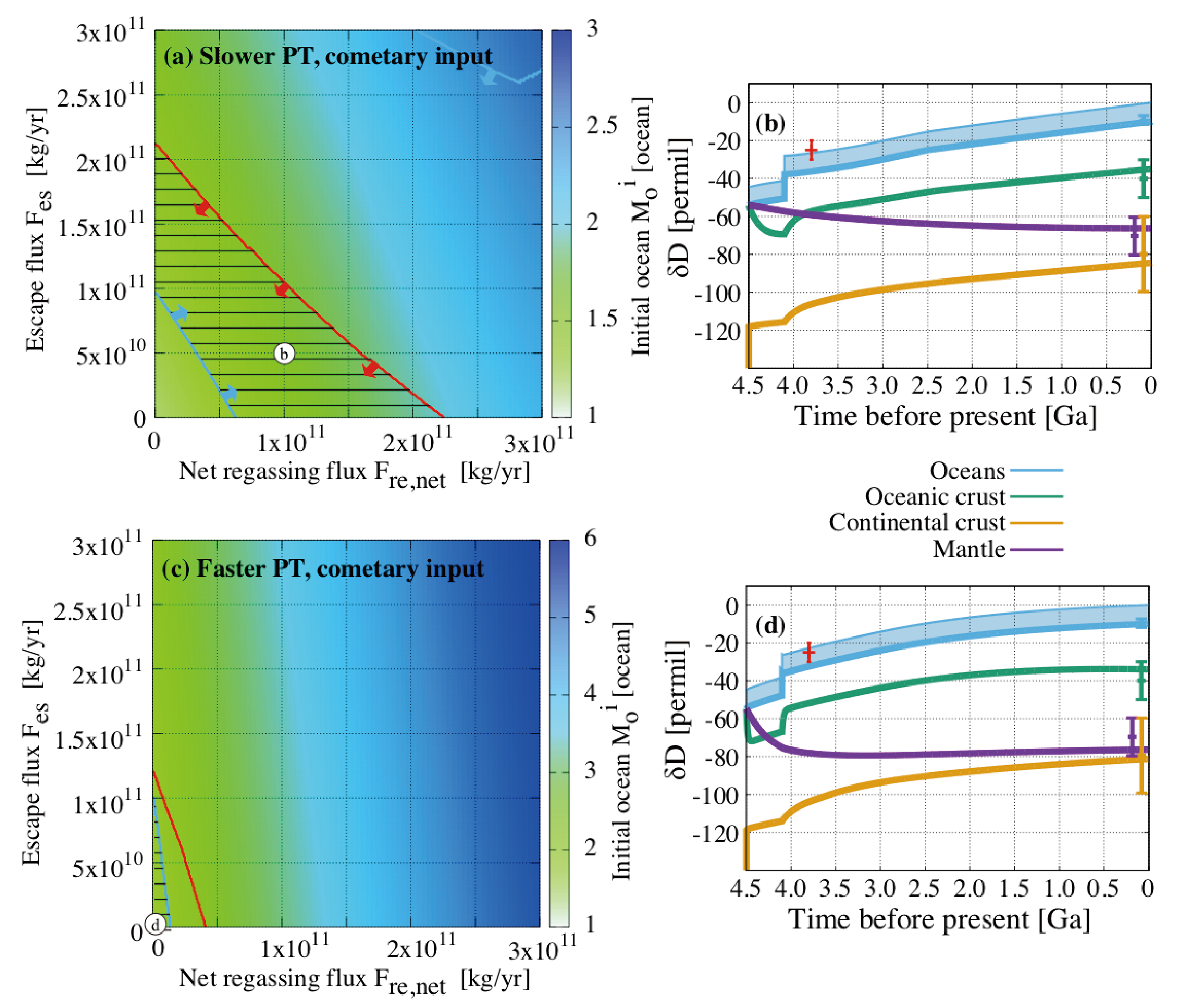}
    \caption{The same as Figure \ref{fig:FreFes}, but cometary input was assumed at $4.1$ Ga (see text). 
    }
    \label{fig:comets}
\end{figure}

The ranges of $F_{\rm re,net}$ and $F_{\rm es}$ where the constraints on present-day and Archean D/H were satisfied are shown for the slower and faster PT models in Figures \ref{fig:FreFes}a and \ref{fig:FreFes}b, respectively.
The standard model was assumed for fractionation factors.
The initial water mass in the oceans $M_{\rm o}^{\rm i}$ is also shown.
We calculated the evolution of masses and D/H ratios of the water reservoirs for each set of $F_{\rm re,net}$ and $F_{\rm es}$.
Examples of evolutionary tracks are shown in Figures \ref{fig:Ts} and \ref{fig:Tf}.
The results for different values of fractionation factors are shown in Figure \ref{fig:FreFes_fdehyfde}.
The influence of the possible late accretion of comets are considered in Figure \ref{fig:comets}.
As explained below, the results showed that the slower PT model assuming $F_{\rm re,net}=0$ and $F_{\rm es}=0$ does not satisfy the constraints on D/H (Figure \ref{fig:FreFes}a).
Therefore, hydrogen escape from the reduced early atmosphere, secular net regassing (Figure \ref{fig:FreFes}a), or faster plate tectonics on early Earth (Figure \ref{fig:FreFes}b) is required.

\subsection{The slower PT model}
\label{results:slower}

The evolution of ${\rm \delta D_i}$ and $M_{\rm i}$ (where i is an arbitrary reservoir) in the slower PT model assuming $F_{\rm re,net}=0$ and $F_{\rm es}=0$ is shown in Figure \ref{fig:Ts}a.
The oceanic ${\rm \delta D_o}$ increased through time because of the isotopic fractionation from seafloor alteration, slab dehydration, and chemical alteration of continents.
All of these processes led to D-enrichment in liquid water.
The water in the oceans $M_{\rm o}$ decreased in response to the increase of the water in sediments $M_{\rm cc}$ because continental growth promoted chemical alteration. 
The mantle ${\rm \delta D_m}$ decreased through time because of the subduction of deuterium-poor water as hydrous minerals.
The continental crust ${\rm \delta D_{cc}}$ increased through time following the increase of the oceanic ${\rm \delta D_o}$.
The oceanic crust ${\rm \delta D_{oc}}$ initially decreased because of the isotopic fractionation by seafloor alteration and then increased in the latter period time following the increase of the oceanic ${\rm \delta D_o}$.
The integrated increase in ${\rm \delta D_o}$ caused by the deep water cycle and continental growth reached $\sim +20$\textperthousand, which is enough to reproduce the increase of ${\rm \delta D_o}$ from the low ${\rm \delta D}$ Archean seawater \citep{Pope+2012}.
However, the present-day ${\rm \delta D_m}$ in the model disagreed with the value inferred from sample analyses \citep{Kyser+ONeil1984,Clog+2013}.
The deep water cycle in our model evolved toward a steady state given by ${\rm \Delta D_{o-m}} \equiv {\rm \delta D_o}-{\rm \delta D_m} \sim 70$\textperthousand, which is consistent with ${\rm \Delta D_{o-m}}$ of the present-day Earth (\ref{appendix-sec1} and \ref{appendix-sec2}), but the rates of degassing and regassing are too small to reach the steady state within 4.5 Gyr.
The constraints on the continental crust ${\rm \delta D_{cc}}$ and oceanic crust ${\rm \delta D_{oc}}$ were satisfied in all cases because both ${\rm \delta D_{cc}}$ and ${\rm \delta D_{oc}}$ are in a steady state with the oceanic ${\rm \delta D_{o}}$ (\ref{appendix-sec1}).

Secular net regassing and hydrogen escape can increase ${\rm \Delta D_{o-m}}$ (Figures \ref{fig:Ts}b and \ref{fig:Ts}c).
The secular regassing transported deuterium-poor water from the oceans to mantle and the hydrogen escape removed deuterium-poor water from the oceans (more precisely, from the atmosphere, which exchanges water with the oceans).
The net regassing on present-day Earth in the slower PT model led to steady decrease and increase in $M_{\rm o}$ and $M_{\rm m}$, respectively (Figures \ref{fig:Ts}b).
The mantle ${\rm \delta D_m}$ started to increase at $\sim 3.5$ Ga following the increase of the oceanic ${\rm \delta D_o}$.
The hydrogen escape removed water from the oceans $M_{\rm o}$ before the GOE at 2.5 Ga (Figures \ref{fig:Ts}c).
In contrast to the secular regassing (Figures \ref{fig:Ts}b), the hydrogen escape and its cessation at 2.5 Ga resulted in the kink in the evolution of ${\rm \delta D_o}$ (and consequently, in that of ${\rm \delta D_{cc}}$ and ${\rm \delta D_{oc}}$) at 2.5 Ga.
The coexistence of the secular regassing diminishes the kink caused by the cessation of hydrogen escape, depending on $F_{\rm re,net}$ and $F_{\rm es}$ (Figures \ref{fig:Ts}d).

A comparison of the results in the slower PT models with the D/H constraints allowed us to constrain $F_{\rm re,net}$ and $F_{\rm es}$ (Figure \ref{fig:FreFes}a).
Because the change in ${\rm \delta D}$ values was too small to reproduce present-day ${\rm \Delta D_{o-m}}$ in the model assuming $F_{\rm re,net}=0$ and $F_{\rm es}=0$ (Figure \ref{fig:Ts}a), this provided a lower limit on $F_{\rm re,net}$ and $F_{\rm es}$ that can satisfy the constraint on present-day D/H (the sky-blue line in Figure \ref{fig:FreFes}a).
We note that there was also a upper limit in these values above which ${\rm \Delta D_{o-m}}$ was too large as seen in Figure \ref{fig:FreFes}b, but it was outside the range of Figure \ref{fig:FreFes}a.
On the other hand, the increase in the oceanic ${\rm \delta D_o}$ inferred from the low ${\rm \delta D}$ of the Archean seawater \citep{Pope+2012} was reproduced in the model assuming $F_{\rm re,net}=0$ and $F_{\rm es}=0$ (Figure \ref{fig:Ts}a).
Because both the secular regassing and hydrogen escape promote the increase in ${\rm \delta D_o}$, there was an upper limit on $F_{\rm re,net}$ and $F_{\rm es}$ to satisfy the constraint on the Archean seawater (the red line in Figure \ref{fig:FreFes}a).
All the constraints on D/H were satisfied in the limited range of $F_{\rm re,net}$ and $F_{\rm es}$ (the hatched area in Figure \ref{fig:FreFes}a).

The range of $F_{\rm re,net}$ and $F_{\rm es}$ depends on the assumed values of fractionation factors.
Within the range of uncertainties (Table \ref{tab:model}), assuming smaller fractionation for dehydration led to the hatched area moving to the right (Figure \ref{fig:FreFes_fdehyfde}a), whereas assuming larger fractionation for degassing led to the hatched area moving to the left (Figure \ref{fig:FreFes_fdehyfde}c).

\textcolor{black}{
The possible late accretion of comets moderately influences the resulting range of $F_{\rm re,net}$ and $F_{\rm es}$, but the D/H constrains still require at least either one of the two mechanisms: secular net regassing or hydrogen escape (Figure \ref{fig:comets}a).
Because comets have a high D/H ratio, the input resulted in an increase in ${\rm \delta D_o}$ (Figure \ref{fig:comets}b), while its contribution to Earth's water budget ($\sim 0.01$ ocean water, Supplementary text S5) is negligible.
The input of D-enriched water decreased $F_{\rm re,net}$ and $F_{\rm es}$ required to reproduce the present-day D/H ratios and increased these fluxes required to reproduce the D/H ratio of the Archean seawater.
}

\subsection{The faster PT model}
\label{results:faster}

In contrast to the slower PT model assuming $F_{\rm re,net}=0$ and $F_{\rm es}=0$ (Figure \ref{fig:Ts}a), the efficient water cycle on early Earth in the faster PT model led to the ${\rm \delta D}$ values nearly reaching steady state in ${\rm \Delta D_{o-m}}$ (\ref{appendix-sec1}) even assuming $F_{\rm re,net}=0$ and $F_{\rm es}=0$ (Figure \ref{fig:Tf}a).
As with the slower PT model, the isotopic fractionation resulted from the seafloor alteration, slab dehydration, and chemical alteration of continents. 
The faster PT model assuming the secular regassing showed a continuous increase in both the oceanic ${\rm \delta D_o}$ and mantle ${\rm \delta D_m}$ (Figure \ref{fig:Tf}b), which is qualitatively the same as the slower PT model (Figure \ref{fig:Ts}b), but the change is larger because of the higher regassing rate in the past.
Though the hydrogen escape had a minor influence on the evolution of the water masses in the reservoirs, its effect on the evolution of D/H was more pronounced (Figures \ref{fig:Tf}c and \ref{fig:Tf}d) because of the efficient fractionation by hydrogen escape compared to the other processes (Table \ref{tab:model}).

The faster PT models assuming $F_{\rm re,net}=0$ and $F_{\rm es}=0$ satisfied both the constraints from D/H of the present-day water reservoirs and of the Archean seawater (Figure \ref{fig:FreFes}b).
Because assuming the secular regassing or hydrogen escape increased the change in ${\rm \delta D_o}$ and the present-day ${\rm \Delta D_{o-m}}$ (Figure \ref{fig:Tf}), both the constraints on present-day and Archean D/H gave an upper limit on $F_{\rm re,net}$ and $F_{\rm es}$ (the purple and red lines in Figure \ref{fig:FreFes}b, respectively).
Neither the secular regassing nor hydrogen escape is necessarily required and the allowed range of $F_{\rm re,net}$ and $F_{\rm es}$ was smaller than that in the slower PT model.

Changing the values of fractionation factors in the faster PT model showed a similar behavior to the slower PT model (Figures \ref{fig:FreFes_fdehyfde}b and \ref{fig:FreFes_fdehyfde}d), but the key result---neither the secular regassing nor hydrogen escape is necessarily required in the faster PT model---does not change.

\textcolor{black}{
Assuming cometary input slightly changed the range of $F_{\rm re,net}$ and $F_{\rm es}$ allowed to reproduce D/H constraints (Figures \ref{fig:comets}c).
The input of D-enriched water (Figures \ref{fig:comets}d) decreased $F_{\rm re,net}$ and $F_{\rm es}$ allowed to reproduce the present-day D/H ratios and increased these fluxes allowed to reproduce the D/H ratio of Archean seawater.
Again, the results showed that neither the secular regassing nor hydrogen escape is necessarily required in the faster PT model.
}

Compared to the slower PT models (Figures \ref{fig:FreFes}a and \ref{fig:Ts}), the faster PT models resulted in a large decrease in the oceanic volume (Figures \ref{fig:FreFes}b and \ref{fig:Tf}).
In the faster PT models, different dependence on the thermal evolution is assumed for the regassing (Equation \ref{eq:Fre}) and degassing (Equation \ref{eq:Fde}) as predicted by the boundary-layer model of the thermal evolution \citep{Honing+2016}.
The difference resulted in the net regassing being larger in the earlier period.
Even in the case where the balance was assumed for the present-day Earth ($F_{\rm re,net}=0$), $\sim$0.7 oceans of water subducted in 4.5 Gyr (Figure \ref{fig:Tf}a).
The faster PT models assuming the net regassing today showed much higher regassing in the past, leading to $\sim 4.3$ oceans of water subducted throughout Earth's history (Figure \ref{fig:Tf}b).

\section{Discussion}
\label{Discussion}

\subsection{Evolution of water on Earth constrained by D/H}
\label{discussion:evo}

The evolution of water on Earth can be constrained by comparing our results with the D/H ratios of present-day water reservoirs (subsection \ref{model:D/H} and Table \ref{Table_reservoirs}).
The differences in ${\rm \delta D}$ between the present-day oceans, continental and oceanic crust, and mantle were shown to result from isotopic fractionation through the seafloor alteration, slab dehydration, and chemical alteration (Section \ref{Results}).
The D/H ratios of present-day reservoirs can be understood by using the steady state (\ref{appendix-sec1} and \ref{appendix-sec2}).
We note that a steady state in hydrogen isotope compositions of the oceans and mantle has been proposed by previous studies \citep{Taylor1974,Javoy2005}, though dehydration-induced fractionation has not been considered in them.
However, the model also showed that the rates of present-day regassing and degassing are small so that the system does not reach the present-day ${\rm \Delta D_{o-m}}$ within 4.5 Gyr (Figure \ref{fig:Ts}a).
Therefore, the hydrogen escape from the reduced early atmosphere, secular regassing, or faster plate tectonics on early Earth was needed to explain the present-day state of D/H ratios in the reservoirs (Figure \ref{fig:FreFes}).

The low ${\rm \delta D}$ of Archean seawater \citep{Pope+2012} further constrains the evolution of water on Earth (subsection \ref{model:D/H}).
Both the slower and faster PT models assuming $F_{\rm re,net}=0$ and $F_{\rm es}=0$ resulted in the secular increase in the oceanic ${\rm \delta D_o}$ as a result of the deep water cycle and continental growth, which agreed with the constraint on the Archean seawater (Figures \ref{fig:Ts}a and \ref{fig:Tf}a).
Because the secular regassing and hydrogen escape promote the increase in ${\rm \delta D_o}$, the constraint on ${\rm \delta D}$ of the Archean seawater gave an upper limit on $F_{\rm re,net}$ and $F_{\rm es}$ (Figure \ref{fig:FreFes}).

These three possibilities---the hydrogen escape, faster plate tectonics on early Earth, and secular regassing---are mutually exclusive.
For instance, in our standard model, the slower PT model needed $F_{\rm re,net}= 0.6-2.1 \times 10^{11}$ kg/yr or $F_{\rm es} = 0-1.2 \times 10^{11}$ kg/yr (Figure \ref{fig:FreFes}a).
Assuming larger $F_{\rm re,net}$ leads to lower $F_{\rm es}$, and vice versa.
On the other hand, the faster PT model allowed much smaller values: $F_{\rm re,net}= 0-0.2 \times 10^{11}$ kg/yr and $F_{\rm es} = 0-1.0 \times 10^{11}$ kg/yr.

Considering the possible ranges of fractionation factors and \textcolor{black}{late accretion of comets (Figures \ref{fig:FreFes}, \ref{fig:FreFes_fdehyfde}, and \ref{fig:comets})} yielded $F_{\rm re,net}< 3.9 \times 10^{11}$ kg/yr and \textcolor{black}{$F_{\rm es}< 2.1 \times 10^{11}$ kg/yr}.
The upper limit of the present-day regassing rate is higher than the upper limit proposed by \citet{Parai+Mukhopadhyay2012} from the constraint on the sea-level change ($F_{\rm re,net}= 1.0 \times 10^{11}$ kg/yr), but is comparable with the range recently argued by \citet{Korenaga+2017} considering the change in \textcolor{black}{buoyancy of continental lithosphere relative to oceanic lithosphere} through time: $F_{\rm re,net}= 3.0-4.5$ $\times 10^{11}$ kg/yr).
The upper limit of the rate of water loss due to hydrogen escape \textcolor{black}{$F_{\rm es} = 2.1 \times 10^{11}$ kg/yr} corresponds to \textcolor{black}{$\sim$900 ppmv} of CH$_4$ (Equation S1).
The high CH$_4$ concentration \textcolor{black}{might} be possible during the early Archean \citep{kharecha+2005}.

The secular regassing, hydrogen escape, and chemical alteration promoted by continental growth contributed to the secular decrease of oceanic water in our model.
In the slower PT model, the possible maximum value of $M_{\rm o}^{\rm i}$ is given when $F_{\rm re,net}= 3.9 \times 10^{11}$ kg/yr was assumed (Figure \ref{fig:FreFes_fdehyfde}a).
The initial mass of the bulk oceans was 2.5 oceans (considering 1.3 oceans later subducted and 0.2 oceans formed sediments). 
In the faster PT model, the possible maximum value of $M_{\rm o}^{\rm i}$ is given when $F_{\rm re,net}= 0.83 \times 10^{11}$ kg/yr was assumed (Figure \ref{fig:FreFes_fdehyfde}b), the initial mass was 2.9 oceans (1.7 oceans subducted and 0.2 oceans formed sediments).
These values are comparable with the estimate of \citet{Korenaga+2017} to reconcile the continental freeboard with the geologic constraints. 

We suggest that the D/H constraints are consistent with the secular regassing scenario where water in the present-day mantle entirely resulted from the regassing through out Earth's history.
The scenario is consistent not only with the geologic constraints on continental freeboard when the change in relative buoyancy of continental \textcolor{black}{lithosphere} is taken into account \citep{Korenaga+2017}, but also with the theoretical predictions of crystallization of magma oceans, where the majority of water was partitioned into the atmosphere and oceans \citep{Hamano+2013}.
Such initial conditions have been argued to be ideal for initiating the plate tectonics \citep{Korenaga2013}. 

\subsection{Implications for future analysis of D/H on early Earth}

\begin{figure}
    \centering
    \includegraphics[width=14cm]{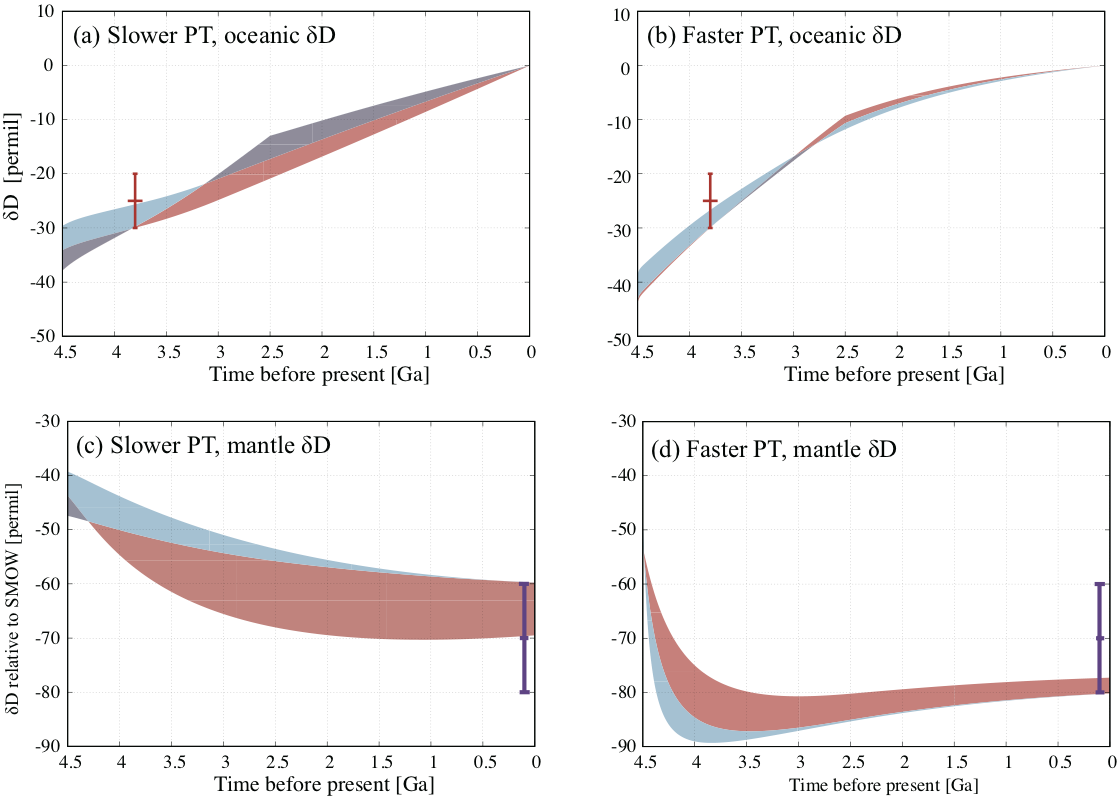}
    \caption{${\rm \delta D}$ of (a,b) oceans and (c,d) mantle as a function of time. The results for (a,c) the slower and (b,d) faster PT models are shown. Shaded areas show the possible ranges of evolutionary tracks that correspond to the parameter sets in hatched areas in (a,c) Figure \ref{fig:FreFes}a and (b,d) Figure \ref{fig:FreFes}b along the sky-blue and red lines. Colors of the shaded ranges correspond to those of boundary lines in Figure \ref{fig:FreFes}. Data points are ${\rm \delta D}$ of Archean seawater and of present-day mantle (the same as Figure \ref{fig:Ts}).}
    \label{fig:t-dDom}
\end{figure}

The three scenarios that explain the ${\rm \delta D}$ values of the present-day and Archean reservoirs---the hydrogen escape, secular regassing, and faster plate tectonics on early Earth---can be distinguished by future analyses of samples that record the D/H ratios of the seawater {\it and mantle} on early Earth.
The evolution of the oceanic and mantle ${\rm \delta D}$ for various sets of $F_{\rm re,net}$ and $F_{\rm es}$ and for the slower and faster PT models is shown in Figure \ref{fig:t-dDom}.
Though the increase in the oceanic ${\rm \delta D_o}$ has been interpreted as a signature of hydrogen escape \citep{Pope+2012}, Figures \ref{fig:t-dDom}a and \ref{fig:t-dDom}b showed that the increase is possible without the escape.
Instead, the kink at the time of the GOE, if confirmed, would be a signature of hydrogen escape from reduced early atmosphere.
Constraining the oceanic ${\rm \delta D_o}$ at the time of the GOE would help us to distinguish the scenarios.

While the difference in the oceanic ${\rm \delta D_o}$ between the slower and faster PT models is small (Figures \ref{fig:t-dDom}a and \ref{fig:t-dDom}b), the mantle ${\rm \delta D_m}$ signal can discriminate between these scenarios (Figures \ref{fig:t-dDom}c and \ref{fig:t-dDom}d).
In contrast to the slower PT model where the mantle ${\rm \delta D_m}$ decreased continuously, the faster PT model showed a rapid change in the earlier period.
Identifying the past mantle ${\rm \delta D_m}$ as well as the oceanic ${\rm \delta D_o}$ would allow us to distinguish the three scenarios.
In addition to the D/H ratios of the oceans and mantle in the Archean, those from the Hadean to the Phanerozoic, if combined with our model, would be useful to constrain the evolutionary scenarios.

\subsection{Comparison with previous studies}

Earth's deep water cycle and loss by hydrogen escape have been investigated in previous studies by using constraints on D/H.
\citet{Lecuyer+1998} modeled the deep water cycle considering two reservoirs: the oceans and mantle. 
They argued that hydrogen isotope variations of the oceans in time may have occurred in response to the imbalance between the rates of regassing and degassing.
Their model showed that the system eventually reached a steady state in ${\rm \delta D_o}$.
In contrast, our model showed that ${\rm \delta D_o}$ kept changing (increasing) in response to the imbalance (regassing) (Section \ref{Results}).
The difference originated from the assumption about the isotopic fractionation: they assumed constant ${\rm \delta D}$ values both for regassed and degassed water, whereas we calculated the ${\rm \delta D}$ values from the fractionation factors (subsection \ref{sub:model}).
In that regard, the results of our model \textcolor{black}{is} more realistic than those of \citet{Lecuyer+1998}.

\citet{Shaw+2008} also modeled the long-term water cycle between the oceans and mantle by taking the isotopic fractionation caused by the seafloor alteration and slab dehydration into account.
Assuming the net balance between the regassing and degassing, they showed the continuous increase and decrease in ${\rm \delta D_o}$ and ${\rm \delta D_m}$, respectively.
Their model even resulted in all the deuterium in the system being partitioned into the oceans after the model was integrated for a long enough time.
Contrary to \citet{Shaw+2008}, our model showed that assuming the balance in the fluxes resulted in ${\rm \delta D_o}$ and ${\rm \delta D_m}$ reaching a steady state determined by the fractionation factors (\ref{appendix-sec2}).
The difference originated from the treatment of fractionation factor of the seafloor alteration: they defined the fractionation factor as the ratio of the altered MORB D/H to unaltered MORB D/H, while we defined it as the ratio of altered MORB D/H to seawater D/H.
Because the water in the hydrous minerals produced by the seafloor alteration originated from seawater, our assumption is more realistic than that of \citet{Shaw+2008}.

\citet{Pope+2012} used mass-balance calculations to derive the amount of water lost by the hydrogen escape based on ${\rm \delta D}$ of Archean seawater.
They concluded that $\sim 0.1$ oceans of water was lost due to the escape.
In contrast, we showed that the lower ${\rm \delta D}$ of Archean seawater can be explained by the isotopic fractionation caused by the deep water cycle (seafloor alteration and slab dehydration) and that the hydrogen escape is not necessarily needed (Section \ref{Model}).
The conclusions differ because they neglected the influence of the regassing and degassing on the ${\rm \delta D}$ change in the mass-balance calculations.
They justified neglecting the contribution of the mantle to the evolution of ${\rm \delta D_o}$ by assuming the net balance between the regassing and degassing.
However, ${\rm \delta D_o}$ changes because of the deep water cycle, even in the case where the net balance was assumed (\ref{appendix-sec2}).

\subsection{Other data sets of Precambrian seawater D/H}
\label{ss:datasets}

\textcolor{black}{
We adopted the D/H ratio of Archean seawater at 3.8 Ga constrained by \citet{Pope+2012}, but other studies have also estimated the paleo-seawater D/H ratio \citep{Lecuyer+1996,Kyser+1999,Hren+2009}.
Whereas \citet{Pope+2012} has reported ${\rm \delta D} = -25 \pm 5$\textperthousand \ and ${\rm \delta ^{18} O} = 2.3$\textperthousand \ from serpentine samples from 3.8 Ga Isua Supracrustal Belt in West Greenland, \citet{Hren+2009} has constrained ${\rm \delta D} = -70$\textperthousand \ to $-5$\textperthousand \ and ${\rm \delta ^{18} O} = -18$\textperthousand \ to $-8$\textperthousand \ from 3.4 Ga Buck Reef Chert rocks in South Africa.
Also, by assuming ${\rm \delta ^{18} O} = -10$\textperthousand \ as suggested by a theoretical model \citep{Kasting+2006}, \citet{Hren+2009} has proposed that the Archean seawater had ${\rm \delta D} = -60$\textperthousand \ .
Reproducing both ${\rm \delta D} = -25 \pm 5$\textperthousand \ at 3.8 Ga and ${\rm \delta D} = -60$\textperthousand \ at 3.4 Ga in our model is difficult and might require a singular event between the interval. 
The discrepancy in ${\rm \delta ^{18} O}$ values in the two studies suggests that these data might be incompatible.
Because chert has lower concentrations of water than serpentine and the temperature dependence of D/H fractionation between chert and water is not straightforward, \citet{Pope+2012} argued that chert is a less well-suited proxy, so that we chose to use \citet{Pope+2012}'s data as a reliable anchor in the Archean for the evolution of seawater D/H.
}

\textcolor{black}{
\citet{Lecuyer+1996} has proposed ${\rm \delta D} = 0 \pm 20$\textperthousand \ from mafic-ultramafic samples from the Chukotat Group of the Lower Proterozoic (2.0-1.9 Ga) Cape Smith fold belt.
They have reported the ${\rm \delta D}$ values matching closely those found in modern metavolcanic rocks.
\citet{Kyser+1999} has constrained the D/H ratio of 2.8-2.6 Ga seawater as ${\rm \delta D} > -20$\textperthousand \ from serpentine minerals from Archean Abitibi greenstone belt in Ontario.
Because the estimates of these earlier studies were less systematic compared to \citet{Pope+2012}, we did not plot them explicitly.
However, these estimates are in accordance with the trend from the ${\rm \delta D}$ value of \citet{Pope+2012} to that of present-day seawater so that adopting them would hardly change our results quantitatively.
}

\subsection{Subduction regimes}

This study investigated the secular evolution of the global water cycle with the numerical modeling under a wide range of parameter spaces (Figure \ref{fig:FreFes}), yet assumed a limited case of cold subduction (Supplementary text S3, $10^3 {\rm ln} f_{\rm dehy}=-40$ to $-23$) where the major water release from the slab to the surface ($F_{\rm ar}$) occurs mainly due to dehydration of the hydrous minerals of amphibole and lawsonite \citep{Schmidt+Poli1998,Maruyama+Okamoto2007}. 
However, our model also explored two extreme subduction regimes of faster and slower PT. 
Subduction velocity along with plate thermal structure and wedge mantle viscosity strongly influences slab surface temperature, which defines the stability of hydrous minerals \citep{Kincaid+Sacks1997,Peacock1993}.
Each hydrous mineral may have a different fractionation factor of D/H due to decomposition. 
In the case of a hot slab, water released by major dehydration events would be in isotopic equilibrium with amphibole ($\sim$350$^\circ$ C), epidote and amphibole ($\sim$400$^\circ$ C), and chlorite, epidote, and amphibole ($\sim$550$^\circ$ C), respectively \citep{Maruyama+Okamoto2007}.
The fractionation factors $f_{\rm dehy}$ of these hydrous minerals range from $10^3 {\rm ln} f_{\rm dehy} =-23$ to $-48$ \citep{Suzuoki+Epstein1976,Graham+1984,Chacko+1999}.
As this range is similar to that in the case of a cold slab, our results would also apply to the case of a hot slab, though the regassing would mainly be contributed to by the cold-slab subduction.
The faster PT might have a smaller net regassing flux $F_{\rm re}$ because of the dominance of hot slabs, but our model acknowledged this and involved such uncertainties of $F_{\rm re}$.

\subsection{Onset time of plate tectonics}

Our model assumed that plate tectonics has operated since 4.5 Ga, soon after the solidification of magma oceans.
The onset time of plate tectonics is controversial, but some studies have suggested that, on the basis of the geochemistry of Hadean zircons, plate tectonics may have already been operating in the Hadean \citep[][and references therein]{Harrison2009,Korenaga2013}.
\citet{Korenaga2013} argued that the ideal initial water distribution to drive plate tectonics is voluminious oceans underlain by a dry mantle.
Such conditions were shown to be consistent with the constraints from hydrogen isotopes (subsection \ref{discussion:evo}), though our model can apply to other onset times of plate tectonics.

\subsection{Uncertainty in the continental growth model}
\label{ss:continents}

\textcolor{black}{The behavior of our model depended only weakly on the assumed continental growth model.  Thus we demonstrate that the evolution of D/H and surface water mass is a poor constraint on continent formation.
As sensitivity analysis, we compared the results using the continental growth models of \citet{Mclennan+Taylor1982} (Figure \ref{fig:FreFes}) and \citet{Armstrong1981} (Figure S1) where the continental coverage increased linearly in the first 1 Gyr and kept a constant value.
We found that the dependence of the results on the uncertainty in the continental growth model is small compared to that on the uncertainty in the fractionation factors (Figure \ref{fig:FreFes_fdehyfde}).
}

\section{Conclusions}
\label{Conclusion}

We modeled the evolution of the masses and D/H ratios of water in the oceans, continental and oceanic crust, and mantle.
The model considered water transport and hydrogen isotopic fractionation by seafloor hydrothermal alteration, chemical alteration of continental crust, slab subduction, hydrogen escape, and degassing at mid-ocean ridges, hot spots, and arcs.
The differences in D/H ratios between the present-day oceans, oceanic and continental crust, and mantle were shown to result from isotopic fractionation by seafloor alteration, slab dehydration, and chemical weathering.
The current degassing and regassing rates were too small to reach the present-day D/H, so an additional mechanism was required.
We showed three evolutionary scenarios that can account for the present-day D/H ratios: (a) hydrogen escape from a reduced early atmosphere, (b) secular net regassing, and (c) faster plate tectonics on early Earth expected from conventional thermal evolution models.
A low D/H ratio of Archean seawater at 3.8 Ga has been interpreted as a signature of the hydrogen escape from a reduced early atmosphere.
However, our model showed that the secular net regassing through out Earth's history or faster plate tectonics on early Earth can also reproduce the constraints on D/H.
These three scenarios are mutually exclusive. 
The rates of hydrogen escape from early Earth and secular regassing on present-day Earth are constrained to be lower than \textcolor{black}{$2.1\times 10^{11}$ kg/yr} and $3.9\times 10^{11}$ kg/yr, respectively.
The initial oceans could be 2--3 times as voluminous as that on current Earth in the secular regassing scenario.
A signature of hydrogen escape before the GOE is visible as a kink in the slope of the evolution of the oceanic D/H ratio.
The mantle D/H ratio in the faster plate tectonics model decreased only during the earlier period, whereas the slower plate tectonics model predicted the mantle D/H ratio has been continuously decreasing throughout Earth's history.
\textcolor{black}{Therefore, we emphasize the importance of measurements to constrain mantle D/H ratio throughout Earth's history}, in addition to that of seawater at the time of the GOE, to distinguish these three evolutionary scenarios.

\section*{Acknowledgements}

We thank two anonymous reviewers for comments and suggestions.
HK was supported by JSPS KAKENHI Grant (15J09448) and JSPS Core-to-Core Program \textquotedblleft International Network of Planetary Sciences.\textquotedblright \
CH was primarily supported by the WPI-funded Earth-Life Science Institute at Tokyo Institute of Technology as well as additional support through MEXT KAKENHI grant (15H05832).
TU was supported by JSPS KAKENHI Grant (17H06454,17H06459).

\newpage

\appendix

\section{Derivation of a steady state in D/H}
\label{appendix-sec1}

The evolution of ${\rm \delta D}$ values of reservoirs can be understood as the change toward a steady state.
From Equation \ref{eq:MiIi}, the steady state is given by,
\begin{equation}
    \frac{I_{\rm o}}{I_{\rm m}} = \frac{F_{\rm de}f_{\rm de}}{F_{\rm re}f_{\rm re}} \frac{F_{\rm re}f_{\rm re}+F_{\rm ar}f_{\rm ar}}{F_{\rm se}f_{\rm se}+F_{\rm ch}f_{\rm ch}} \label{eq:IoIm}
\end{equation}
\begin{equation}
    \frac{I_{\rm cc}}{I_{\rm o}} = \frac{F_{\rm ch}f_{\rm ch}}{F_{\rm we}f_{\rm we}} \label{eq:IccIs}
\end{equation}
\begin{equation}
    \frac{I_{\rm oc}}{I_{\rm o}} = \frac{F_{\rm se}f_{\rm se}+F_{\rm ch}f_{\rm ch}}{F_{\rm re}f_{\rm re}+F_{\rm ar}f_{\rm ar}}. \label{eq:IocIs}
\end{equation}
For the parameter values used in our study (Table \ref{tab:model}) where water in oceanic crust mostly degassed through arc volcanism, $F_{\rm ar} \gg F_{\rm re}$, $F_{\rm se} \gg F_{\rm ch}$, and $F_{\rm ch} \sim F_{\rm we}$. 
Here we further assume $F_{\rm de} = F_{\rm re}$. 
Equations \ref{eq:fdehy1} and \ref{eq:fdehy2} are approximated as $f_{\rm re} \sim f_{\rm dehy}$ and $f_{\rm ar} \sim 1$.
Substituting these relations for Equations \ref{eq:IoIm}-\ref{eq:IocIs} gave,
\begin{equation}
    \frac{I_{\rm o}}{I_{\rm m}} \sim \frac{f_{\rm de}}{f_{\rm se}f_{\rm dehy}} \label{eq:IoIm_d}
\end{equation}
\begin{equation}
    \frac{I_{\rm cc}}{I_{\rm o}} \sim \frac{f_{\rm ch}}{f_{\rm we}} \label{eq:IccIs_d}
\end{equation}
\begin{equation}
    \frac{I_{\rm oc}}{I_{\rm o}} \sim f_{\rm se}. \label{eq:IocIs_d}
\end{equation}
Therefore, we can derive,
\begin{equation}
    {\rm \Delta D_{o-m}} \sim 10^3 {\rm ln} f_{\rm de} - 10^3 {\rm ln} f_{\rm se} -10^3 {\rm ln} f_{\rm dehy} \label{eq:dDo-m_d}
\end{equation}
\begin{equation}
    {\rm \Delta D_{cc-o}} \sim 10^3 {\rm ln} f_{\rm ch} - 10^3 {\rm ln} f_{\rm we} \label{eq:dDcc-o_d}
\end{equation}
\begin{equation}
    {\rm \Delta D_{oc-o}} \sim 10^3 {\rm ln} f_{\rm se} \label{eq:dDoc-o_d}
\end{equation}
Substituting the fractionation factors (Table \ref{tab:model}) gave ${\rm \Delta D_{o'-m}} \sim 70$\textperthousand, ${\rm \Delta D_{cc-o'}} \sim -80$\textperthousand, and ${\rm \Delta D_{oc-o'}} \sim -30$\textperthousand, which is consistent with the ${\rm \delta D}$ values obtained from field samples (Table \ref{Table_reservoirs}).
Here an apostrophe denotes the fractionation from seawater before the correction by adding small reservoirs.

On the other hand, considering a water cycle without slab dehydration ($F_{\rm re} \gg F_{\rm ar}$) leads to a different steady state.
In this case Equation \ref{eq:IoIm} gave,
\begin{equation}
    \frac{I_{\rm o}}{I_{\rm m}} \sim \frac{f_{\rm de}}{f_{\rm se}} \label{eq:IoIm_n}
\end{equation}
\begin{equation}
    {\rm \Delta D_{o-m}} \sim 10^3 {\rm ln} f_{\rm de} - 10^3 {\rm ln} f_{\rm se}. \label{eq:dDo-m_n}
\end{equation}
Equation \ref{eq:dDo-m_n} led to ${\rm \Delta D_{o'-m}} \sim 30$\textperthousand, which is smaller than the measurements (Table \ref{Table_reservoirs}).
These estimates showed that the isotopic fractionation due to slab dehydration is one of the important processes to explain the present-day D/H ratios in the mantle.

\section{The constant-flux model}
\label{appendix-sec2}

\begin{figure}
    \centering
    \includegraphics[width=12cm]{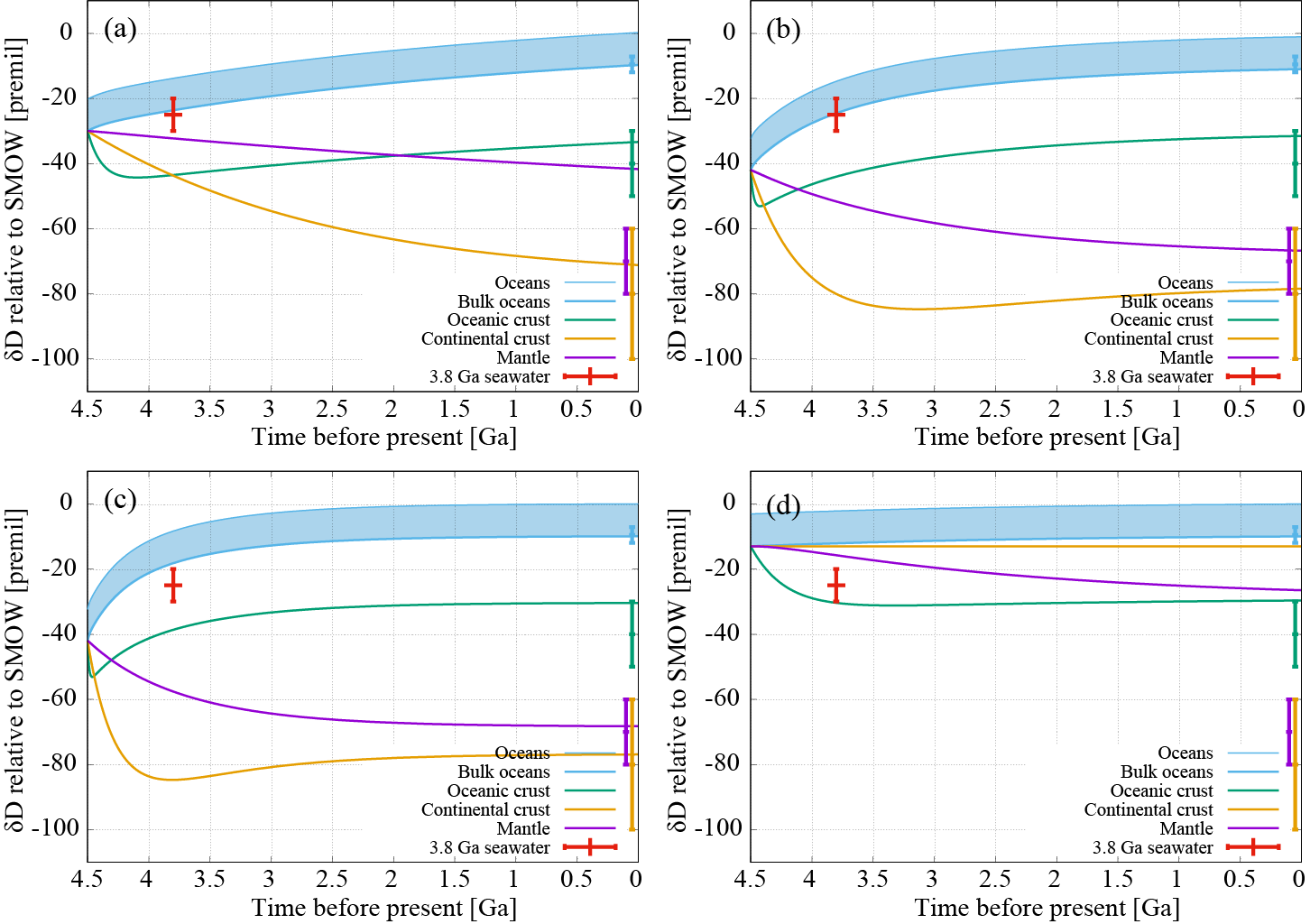}
    \caption{${\rm \delta D}$ of oceans (thin cyan lines), bulk oceans (thick cyan lines), oceanic crust (green lines), continental crust (yellow lines), and mantle (purple lines) as a function of time in the constant-flux model (see text). The shaded range denotes the possible range of oceanic ${\rm \delta D}$ (see subsection \ref{model:D/H}). Results for the dehydration model (a: $F_{\rm de} = 1\times 10^{11}$ kg/yr, b: $F_{\rm de} = 5\times 10^{11}$ kg/yr, and c: $F_{\rm de} = 10\times 10^{11}$ kg/yr) and the non-dehydration model (d: $F_{\rm de} = 5\times 10^{11}$ kg/yr) are shown. Data points are ${\rm \delta D}$ values of reservoirs on present-day Earth (Table \ref{Table_reservoirs}) and 3.8 Ga seawater \citep{Pope+2012}.}
    \label{fig:steady}
\end{figure}

We demonstrated how the ${\rm \delta D}$ evolved toward a steady state (\ref{appendix-sec1}) in Figure \ref{fig:steady}.
Here we assumed that all fluxes are balanced and two types of water cycle were investigated: $F_{\rm de} = F_{\rm ch} = F_{\rm we} = 1/10\ F_{\rm se} = 1/10\ F_{\rm ar}$ (the dehydration model) and $F_{\rm de}  = F_{\rm se} = F_{\rm ar}, F_{\rm ch} = F_{\rm we} = 0$ (the non-dehydration model).
The dehydration model (Figures \ref{fig:steady}a-c) evolved toward a steady state given by Equation \ref{eq:dDo-m_d}, where ${\rm \Delta D_{o'-m}} \sim 70$\textperthousand.
On the other hand, the non-dehydration model (Figure \ref{fig:steady}d) evolved toward a steady state given by Equation \ref{eq:dDo-m_n}, where ${\rm \Delta D_{o'-m}} \sim 30$\textperthousand.

The timescale to reach the steady state is determined by the residence time of water in the ocean and mantle, which is $\sim 10$ Gyr in Figure \ref{fig:steady}a, $\sim 2$ Gyr in Figures \ref{fig:steady}b and \ref{fig:steady}d, and $\sim 1$ Gyr in Figure \ref{fig:steady}c.


\clearpage

\section*{S1: Water budgets and D/H}
\label{budgets}

Estimates for the water masses of four reservoirs (the oceans, oceanic crust, continental crust, and mantle) are summarized in Table 1. 
In our definition, the oceans include small reservoirs that exchange water (and hydrogen) with seawater in short timescales \citep[$< 10^3$ yr,][]{Bodnar+2013}: atmosphere, biosphere, surface water, groundwater, and glaciers/polar ice.
The sum of water in these small reservoirs corresponds to $3\%$ of total water in bulk oceans by mass.
The bulk oceans contain $1.4\ \times 10^{21}\ {\rm kg}$ of water (hereafter referred to as 1 ocean).
The oceanic crust contains 0.10 oceans of water as bound water ($\sim 80\%$) and pore water ($\sim 20\%$) \citep{Jarrard+2003}.
The continental crust contains 0.20 oceans of water mostly as sedimentary rocks ($\sim 87\%$) and metamorphic rocks ($\sim 13\%$) \citep{Bodnar+2013,Korenaga+2017}.

Water content in the mantle is uncertain compared to the other reservoirs.
\citet{Korenaga+2017} estimated the water content in the mantle to be 0.56-1.3 oceans.
\textcolor{black}{Laboratory and computational experiments on natural and synthetic samples} find that the majority of mantle minerals, olivine and pyroxene in the upper mantle and bridgmanite and ferropericlase in the lower mantle, can incorporate around 200 ppm wt. water \citep{Demouchy+2016,Panero+2015}.  \textcolor{black}{Similar values are also observed for the upper mantle source of mid ocean ridge basalts, while more variable and higher values are observed for the source of ocean island basalts.  \citep{Ito+1983,Hirschmann2006}. Thus, the lower bound on the water storage capacity of the upper mantle is around two oceans.}
The transition zone nominally anhydrous minerals wadsleyite and ringwoodite are capable of storing 3.0 wt.\% hydrogen within defects in their crystal structure \citep{Smyth1994,Smyth+2004}.  However, this capacity decreases to around 1.5 wt.\%, or around 5 oceans of water, at transition zone temperatures \citep{Mao+2012}.
Estimates based on seismic observations suggest that the transition zone presently accommodates around one ocean of water \citep{Houser2016}, although localized hydration is possible \citep{Pearson+2014}.
Hydrous silicates that form in subducting oceanic lithosphere are thought to mostly dehydrate by 300 km depth \citep{Green+2010}, but recent studies have found that cold subducting slabs and aluminum can stabilize dense hydrous magnesium silicates such as phase D and phase H allowing hydrogen transport to the deep Earth \citep{Komabayashi2006,Tsuchiya2013,Nishi+2014}.

We assumed $M_{\rm o}^0 = 1$ ocean, $M_{\rm cc}^0 = 0.2$ oceans, and $M_{\rm oc}^0 = 0.1$ oceans.
Most of our models assumed $M_{\rm m}^0 = 1$ ocean, but some models showed integrated regassing larger than 1 ocean.
In these cases, we increased $M_{\rm m}^0$ until the model results in $M_{\rm m}$ at $t = 4.5$ Gyr, which agrees with the assumed $M_{\rm m}^0$.

\section*{S2: Fluxes}
\label{model:flux}

The main part of Earth's deep water cycle is composed of the oceans, oceanic crust, and mantle (Figure 1).
Water is transported by degassing (the mantle to oceans) flux $F_{\rm de}$ at mid-ocean ridges and hot spots, seafloor hydrothermal alteration (the oceans to oceanic crust) flux $F_{\rm se}$, and regassing (the oceanic crust to mantle) flux $F_{\rm re}$ at subduction zones.
On present-day Earth, $F_{\rm de}$ is estimated to be $1.1-3.2\times 10^{11}\ {\rm kg/yr}$ \citep[][and references therein]{Parai+Mukhopadhyay2012,Bodnar+2013}.
The seafloor alteration flux $F_{\rm se}$ is likely to be larger than $F_{\rm de}$ and is estimated as $F_{\rm se} = 9.0-16\times 10^{11}\ {\rm kg/yr}$ \citep[][and references therein]{Lecuyer+1998,Jarrard+2003,Bodnar+2013}.
We note that this value is the minimum because the alteration flux of peridotites in lithospheic mantle, which is excluded in the estimate, might be comparable to that of basalt alteration \citep{Schmidt+Poli1998,Li+Lee2006}.
The majority (up to $\sim 80-90\%$) of the subducted water is expected to return to the oceans directly by updip transport, and indirectly by arc volcanism \citep[through the atmosphere and ground water,][]{Dixon+2002,Bodnar+2013,Korenaga+2017}, but the ratio is controversial \citep{Honing+2016}.

The emergence of the continental crust added two other paths to the global water cycle: chemical alteration of continents and weathering (Figure 1).
The rate at which the surface water (included in the bulk oceans in our model, Table 1) is bound into hydrated mineral phases by chemical alteration $F_{\rm ch}$ is estimated to be $\sim 1\times 10^{11}\ {\rm kg/yr}$ \citep{Bodnar+2013}.
The rate at which water in the form of hydrated minerals in the continental crust eroded and transported to the ocean basins $F_{\rm we}$ is estimated to be $\sim 1\times 10^{11}\ {\rm kg/yr}$ \citep{Bodnar+2013}.
Our model does not explicitly include the water flux transported from the oceanic crust to continental crust by emplacement of hydrous magmas generated in the mantle wedge by dehydration of subducting slab.
This flux is implicitly treated as a part of arc volcanism flux $F_{\rm ar}$ because the input from the oceanic crust by emplacement of hydrous magmas balances the output to bulk oceans (more precisely, to the atmosphere and ground water) by arc magmatism \citep{Bodnar+2013}.
The contribution of pore water in sediments to $F_{\rm se}$ is excluded because the pore water that entered the subduction zone is thought to be entirely expelled from the slab at shallow levels and returned to the oceans \citep{Jarrard+2003}, 
though a sedimentary layer having low permeability is proposed to allow pore water to enter the mantle \citep{Honing+2014,Honing+2016}.

The balance between the degassing $F_{\rm de}$ and regassing $F_{\rm re}$ has been poorly understood.
Geochemical estimates implied $F_{\rm de} < F_{\rm re}$ \citep{Ito+1983}.
Limited sea-level change over the Earth's history inferred from geologic records has been interpreted as suggesting $F_{\rm de} \simeq F_{\rm re}$ \citep[e.g.,][]{Lecuyer+1998,Parai+Mukhopadhyay2012}.
Net regassing flux  $F_{\rm re,net}$ ($\equiv F_{\rm re} - F_{\rm de}$) is estimated to be $< 1\times 10^{11}\ {\rm kg/yr}$ to keep the sea-level change less than 100 m over the Phanerozoic (0.54 Ga to present), as proposed from geologic records \citep{Parai+Mukhopadhyay2012}.
However, \citet{Korenaga+2017} recently argued $F_{\rm re,net} = 3-4.5\times 10^{11}\ {\rm kg/yr}$ (assumed to be constant through time) is needed to keep the sea-level constant by considering that the relative buoyancy of continental lithosphere with respect to oceanic lithosphere was higher in the past.
The estimated net regassing transports 1-1.5 oceans into the mantle, if integrated over the duration of 4.5 Gyr.
We treated the net regassing flux $F_{\rm re,net}$ as a parameter.

Contrary to present-day Earth where hydrogen escape is limited by cold trap of water vapor at tropopause, hydrogen could escape more efficiently because of the transport to the upper atmosphere by reducing species such as CH$_4$.
The rate of hydrogen escape before 2.5 Ga can be approximated by diffusion-limited escape \citep{Kuramoto+2013}.
Assuming the diffusion-limited escape, the rate of water loss due to the hydrogen escape is given by \citep{Catling+2001},
\begin{equation}
    F_{\rm es} = 1.2 \times 10^{14} f_{\rm H}^{\rm tot}\ {\rm kg/yr} \label{eq:Fes-dif}
\end{equation}
where $f_{\rm H}^{\rm tot} \equiv f_{\rm H_2O} + f_{\rm H_2} + 2f_{\rm CH_4}\cdots$ is the total concentration of all H-bearing compounds ($f_{\rm i}$ is the concentration of the species i).
\citet{Pope+2012} proposed $f_{\rm CH_4} =64$ to 480 ppmv as a range consistent with paleosols.
We treated the escape flux $F_{\rm es}$ as a parameter.

We assumed $F_{\rm de}^0 = 1.0 \times 10^{11}\ {\rm kg/yr}$, $F_{\rm se}^0 = 10 \times 10^{11}\ {\rm kg/yr}$, $F_{\rm ch}^0 = 1.5 \times 10^{11}\ {\rm kg/yr}$, $F_{\rm we}^0 = 1.0 \times 10^{11}\ {\rm kg/yr}$, $F_{\rm re}^0 = F_{\rm de}^0+F_{\rm re,net}^0$, and $F_{\rm ar}^0 = F_{\rm se}^0+F_{\rm we}^0-F_{\rm re}^0$, respectively.

\section*{S3: Fractionation factors}
\label{model:f}

Seafloor alteration forms various hydrous minerals and is known to be a process which fractionates D/H. 
The fractionation factors vary among the resulting hydrous minerals as a function of temperature \citep[e.g.,][]{Suzuoki+Epstein1976,Graham+1984,Meheut+2010}.
The averaged value for the seafloor hydrothermal condition is approximately $10^3{\rm ln} f_{\rm se}' = -30$ \citep{Shaw+2008}.
Here an apostrophe denotes the fractionation from the water in oceans before the correction by adding small reservoirs (Section \ref{budgets}).

Fractionation due to slab dehydration has been proposed to contribute to the long-term D/H evolution in the deep water cycle \citep{Shaw+2008,Shaw+2012}.
The details of this fractionation process have been poorly understood.
Here we estimated the fractionation factor by assuming equilibrium fractionation between released water and hydrous minerals.
Studies of phase change of a basalt+H$_2$O system showed that a major dehydration event during subduction is an amphibole-lawsonite transition in the case of a cold slab \citep{Schmidt+Poli1998,Maruyama+Okamoto2007}.
Water released by the transition returns to oceans, while water released by dehydration of lawsonite at much deeper regions is thought to be transported to the mantle \citep{Maruyama+Okamoto2007}.
Whereas the fractionation factor in an amphibole-water system has been reported \citep{Suzuoki+Epstein1976,Graham+1984}, in a lawsonite-water system it has, to the best of our knowledge, not been investigated.
Therefore, we approximated the dehydration-induced fractionation by using the results of experiments for a hornblende-water system: $10^3 {\rm ln} f_{\rm hornblende-water} = -23$ to $-40$ at $300-400\ {\rm ^\circ C}$ \citep{Suzuoki+Epstein1976,Graham+1984}. 
We assumed $10^3{\rm ln} f_{\rm dehy} = -40$ in our standard model.
We note that dehydration of altered peridotites in the lithospheric mantle (lower slab) is not considered to be a fractionation process because this process would resulted in complete dehydration \citep{Maruyama+Okamoto2007}.

Chemical alteration of the continental crust to form hydrous minerals is also known to be a fractionation process of D/H \citep[e.g.,][]{Sheppard1986,Sheppard+Gilg1996,Meheut+2010}.
Analyses of natural samples have shown ${\rm \delta D} = -100$\textperthousand \ to -60\textperthousand \ \citep[Table 1,][]{Lecuyer+1998,Zhou+Dobos1994}.
This could be the result of multiple fractionation processes between oceans-meteoric/surface water and meteoric/surface water-minerals.
We assumed the net fractionation factor $10^3{\rm ln} f_{\rm ch}' = -80$.

Degassing at MORB and at hot spots might fractionate D/H of released water, but the isotopic fractionation would be smaller than those of seafloor hydrothermal alteration and chemical alteration of continents.
OH(l)-OH(v) fractionation has $\Delta {\rm D_{v-l}}\ \sim 20-30$\textperthousand \ at $\sim 950\ {\rm K}$ \citep{Pineau+1998} and $\Delta {\rm D_{v-l}}$ becomes smaller at higher temperature \citep{Dobson+1989}.
Although there are no experiments that measured the fractionation factor at temperatures as high as that of MORB and hot spots \citep[>1700 K,][]{Putirka+2007}, we expect $10^3{\rm ln} f_{\rm de} < 10$.
We assumed $10^3{\rm ln} f_{\rm de} = 0$ in our standard model.

Hydrogen escape due to the photolysis of CH$_4$ produced by photosynthesis and methanogenesis \citep{Catling+2001} was considered in our model.
Methanogen methane has ${\rm \delta D} = -300$\textperthousand \ to -150\textperthousand \ \citep{Catling+2001}.
We assumed $10^3{\rm ln} f_{\rm es} = -150$ to obtain the upper limit of hydrogen escape.

These fractionation factors are corrected in terms of the difference in D/H between the bulk oceans (including small reservoirs) and oceans (excluding small reservoirs).
Assuming a fractionation factor $10^3{\rm ln}f' = 10$ that accounts for the difference, the corrected fractionation factors $f_{\rm k}$ (k denotes an arbitrary process) are written as $f_{\rm k} = f' \times f_{\rm k}'$.

\section*{S4: Initial conditions}

The initial masses and D/H ratios of water would be determined by partitioning and fractionation between reservoirs during the solidification of magma oceans.
The theoretical models assuming equilibrium partitioning of water predicted that more than $\sim 90\%$ of the water would be partitioned into the exosphere at the time of solidification \citep{Hamano+2013,Elkins+2008}. 
The amount of water trapped in the mantle is determined by interstitially trapped melts.
About half of the initial water budget might be trapped in the mantle if degassing becomes inefficient after the transition from the low-viscosity regime to the high-viscosity regime \citep{Hamano+2013}.
Therefore, more than half of the initial water budget should be partitioned into oceans.
The constraint was satisfied in all the models in our calculations.

Because of the high temperature of magma oceans ($>$1500 K), the isotopic fractionation between ocean and mantle (namely, vapor and liquid phases at the time) would be well approximated by equilibrium fractionation.
Whereas fractionations between OH-bearing species (H$_2$O, OH, and MgOH) and others (H$_2$ and H) are relatively large, those among OH-bearing species are less than 10\textperthousand \ at least in the vapor phase \citep{Pahlevan+2016,Pahlevan+2017}.
A residual OH(l)-OH(v) fractionation has $\Delta {\rm D_{v-l}}\ \sim 20-30$\textperthousand \ at $\sim 950\ {\rm K}$ \citep{Pineau+1998} and $\Delta {\rm D_{v-l}}$ becomes smaller at higher temperature \citep{Dobson+1989}.
We assumed that all the reservoirs have the same ${\rm \delta D}$ and iteratively change the value to obtain the present-day oceanic ${\rm \delta D}$ value which equals the SMOW.
We note that carbonaceous chondrites, that are usually considered to be the source of water on Earth, have a range of ${\rm \delta D} = -200$\textperthousand \ to 300\textperthousand \ \citep{Lecuyer+1998}.
We will discuss the cases where we started from different initial conditions of ${\rm \delta D}$ (Appendix C).

\section*{S5: Contribution of the late accretion to water budget and D/H}

\textcolor{black}{
Our model assumed that water had been delivered to Earth before the solidification of magma oceans.
The assumption is supported by the evidence that the mantle late veneer, which gives us a direct clue to late accreting materials, has an inner Solar System origin \citep{Fischer-Godde+Kleine2017} and by the inference that Earth's core likely contains a huge amount of hydrogen \citep{Umemoto+Hirose2015}.}

\textcolor{black}{
While the majority of Earth's water is likely to be delivered early, part of water might have been delivered by the late accretion.
A constraint on the late delivery of volatiles is given by noble gases in the terrestrial atmosphere: the present-day atmosphere contains $22 \pm 5\%$ cometary xenon, which constrains the contribution of comets to Earth's sewawater \citep[$\leq 1\%$,][]{Marty2017}.
The limited contribution of comets to the late accretion is also supported by the noble gas abundances in Martian atmosphere \citep{Kurokawa+2014,Kurokawa+2018}.
The timing of this possible cometary accretion is unknown, but it is usually assumed to be related to the Late Heavy Bombardment, which occurred prior to $\sim 3.8$ Ga \citep{Morbidelli+2012}.
}

\textcolor{black}{
In order to investigate the influence of the possible late accretion of comets, we also explored models where the cometary delivery was implemented by an input of $0.01$ ocean water with $\delta {\rm D} = 1000$\textperthousand \ \citep{Alexander2012+} at $4.1$ Ga.
}

\begin{figure}
    \centering
    \includegraphics[width=10cm]{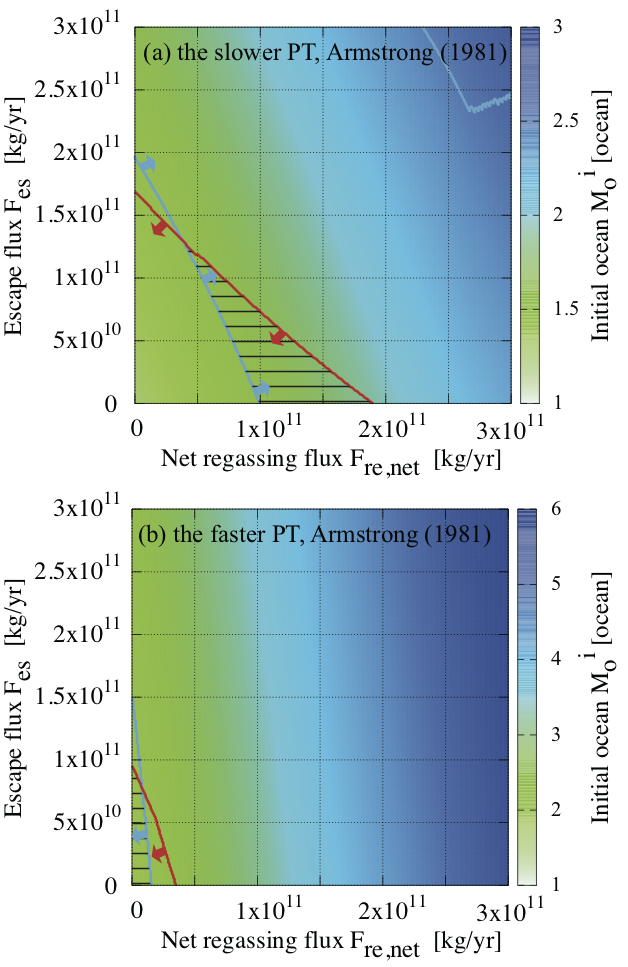}
    \captionsetup{labelformat=empty,labelsep=none}
    \caption{Figure S1: Range of $F_{\rm res,net}$ and $F_{\rm es}$ where the constraints on D/H are satisfied (the hatched areas) in the model where the continental gworth model of \citet{Armstrong1981} was assumed. Results for (a) slower PT model and (b) faster PT model are shown. The present-day D/H ratios of the water reservoirs were reproduced above and below the sky-blue line for (a) and (b), respectively. The D/H of the Archean seawater was reproduced below the red line. Color contour denotes $M_{\rm o}^{\rm i}$. }
    \label{fig:S1}
\end{figure}



 \bibliographystyle{elsarticle-harv}\biboptions{authoryear}

\bibliography{sample}

\end{document}